\documentclass[letterpaper,12pt]{article}

\usepackage{authblk}
\usepackage{graphicx}
\usepackage{braket}
\usepackage{amsmath}
\usepackage{amssymb}
\usepackage{mathtools}
\usepackage{caption}
\usepackage{subcaption}
\usepackage{hyperref}
\usepackage{leftindex}
\usepackage[square,numbers,compress]{natbib}
\usepackage{color}

\usepackage[margin=1in,letterpaper]{geometry}

\providecommand{\keywords}[1]
{
	\small	
	\textbf{\textit{Keywords---}} #1
}

\numberwithin{equation}{section}

\title{\vspace{-1.5cm}\textbf{\large Decoherence Induced by the Noise of Primordial Graviton with Minimum Uncertainty Initial States}}
\author[1]{\normalsize Anom Trenggana \footnote{Electronic address: gstagunganom@gmail.com}}
\author[1,2]{\normalsize Freddy P. Zen \footnote{Electronic address: fpzen@fi.itb.ac.id}}
\author[1]{\normalsize Getbogi Hikmawan \footnote{Electronic address: getbogi@fi.itb.ac.id}}
\affil[1]{\textit{\normalsize Theoretical Physics Laboratory, THEPi (Theoretical High Energy Physics) Division, Faculty of Mathematics and Natural Sciences, Institut Teknologi Bandung, Bandung 40132, West Java, Indonesia}}
\affil[2]{\textit{\normalsize Indonesian Center for Theoritecal and Mathematical Physics (ICTMP), Institut Teknologi Bandung, Bandung 40132, West Java, Indonesia}}
\date{(\today)}

\begin{document}
\maketitle
\begin{abstract}
We have investigated the decoherence induced by the primordial graviton, using the influence functional method, to show whether this method is still effective in detecting graviton if the initial state is not a Bunch-Davies vacuum but rather a minimum uncertainty state. This minimum uncertainty condition allows the initial state of the primordial graviton to be an entanglement state between the polarization or, more generally, a superposition state between a vacuum and that entanglement. Both of those states have a non-classical correlation between the two polarization modes. We found that this method is still effective for detecting gravitons if the density matrix of the initial state does not have non-diagonal elements, where the maximum decoherence time is about 20 seconds, and the dimensions of the interferometer could be reduced if the total graviton increases.
\end{abstract}


\newpage

\tableofcontents

\section{Introduction}\label{introduction}
One prediction from Einstein's general theory of relativity is the phenomenon of gravitational waves \cite{Einstein1,Einstein2}. When there is a compact concentration of energy, spacetime will bend. If the concentration of energy changes, it will produce dynamics in the curvature of spacetime that will propagate in all directions at the speed of light. The propagation of the dynamic curvature of spacetime in all directions is what we know as gravitational waves. In the last decades, gravitational waves have been attempted to be detected. Until 2015, the LIGO detector succeeded in detecting gravitational waves for the first time \cite{LIGO}. This success has become a hope for detecting another phenomenon related to gravitational waves in the future. One of those phenomena is looking for the quantum characteristics of gravitational waves as graviton particles. This phenomenon is interesting because graviton particles are one of the consequences that arise from the quantization of gravity. As we know, there has been no satisfactory theory for the quantum theory of gravity up until now. Some scientists even believe that gravity has not to be quantized at all \cite{Jacobson}. Even so, many scientists are still trying to develop a mechanism to detect gravitons.

Several studies related to graviton detection are still being developed today \cite{PWZ,PWZ2,PWZ3,Kanno3,Kanno4,Anom,Anom2,Maity,Lorenci,Cho}. The interesting one is the research proposed by Parikh \textit{et. al.} \cite{PWZ}. They submitted a study proposal to detect graviton through the quantum noise effect produced by these particles on classical masses with the same vein as quantum Brownian motion \cite{Schwinger,Calderia}. The idea of this detection is based on the assumption that detecting individual gravitons is impossible \cite{Dyson}. In his research, Parikh \textit{et. al.} \cite{PWZ} studied the behavior of a gravitational wave detector in response to a quantized gravitational field. The detector will be modeled as two objects that are geodesically separated. Then, using the Feynman-Vernon influence functional method \cite{Feynman,Feynman2}, the quantum noise generated from the quantized gravitational field can be studied. Furthermore, this study was expanded by S. Kanno \textit{et. al.} \cite{Kanno3}, where, apart from studying the quantum noise correlation of gravitons, in their research, they also learned the decoherence due to the interaction with gravitons in a squeezed coherent state. Another research was conducted by H. T. Cho and B. L. Hu \cite{Cho}, where the study was expanded by considering all possible graviton modes, including polarization. These studies show similar results where quantum noise can be detected if the initial quantum state is squeezed.

Gravitational waves could originate from several places, for example, black holes \cite{Dokuchaev, Hadyan1, Hadyan2}, neutron stars \cite{Hawke, Lasky}, or from the early universe when the inflation mechanism occurred \cite{Watanabe, Boyle, Jinno}. The gravitational wave from the early universe is called a primordial gravitational wave. Some of the best cosmological inflation models that can explain the structure of our universe with great precision predict that the universe started from a quantum state \cite{Starobinsky, Starobinsky2, Guth, Linde, Albrecht, Linde2}. This means that the primordial gravitational waves also originate from the quantum state. So, gravitons would have been generated in the early universe. Interestingly, due to inflation at the beginning of the universe, the initial quantum state was believed to be squeezed as the universe expanded \cite{Vennin}. This means that if gravitons came from the early universe, it should be possible to detect gravitons using the quantum noise method. The proposal to detect primordial gravitons was put forward by S. Kanno \textit{et al.} \cite{Kanno4}. Their research proposed an experimental setup using the entanglement of two macroscopic mirrors suspended at the end of an equal-arm interferometer. If there is an interaction between these mirrors and the primordial graviton, the entanglement state of the mirror will be destroyed. The decoherence time, or the duration of the entanglement between mirrors, will occur for about 20 seconds. The destruction of the entanglement state in 20 seconds can be used as evidence of the existence of the primordial graviton. However, the initial quantum state used is the Bunch-Davies vacuum, which is the initial state that is usually chosen in the conventional inflationary universe. As a result of rapid expansion at the beginning of the universe, the Bunch-Davies vacuum will be squeezed so that in the era of radiation domination, the Bunch-Davies vacuum can be seen as a state of entanglement between the wave number modes $\mathbf{k}$ and $-\mathbf{k}$. Although the initial state is usually chosen to be a vacuum, it does not rule out the possibility that it could take the form of another state. For example, in some research, the initial state can be chosen as $\alpha$-vacua \cite{Kanno8}, namely excited states from the point of the Bunch-Davies vacuum and coherent states \cite{Kanno7}, which can arise in the presence of matter fields during inflation. In this research, we want to expand the understanding of this study by assuming the initial state is a state other than a vacuum. It will be shown whether this method of detecting gravitons is still effective if different initial states are used. Choosing a different initial state should provide a variation in the value of the decoherence time. If the decoherence time lasts very long, then it seems as if the gravitational wave background cannot break the entanglement between the mirrors or there is no decoherence, which means that no evidence can show that there is an interaction with primordial gravitons. For this reason, in this study, we will examine whether there will be a very long decoherence time if the initial state of the primordial graviton is a state other than a vacuum state.

One of the basic concepts that differentiates classical from quantum physics is the Heisenberg uncertainty principle, which is a concept that limits the measurement accuracy of two observables in a system. This limitation exists not because of the inability of the measuring instrument but because of nature itself. The highest measurement accuracy (the condition where measurement accuracy in quantum mechanics is closest to the classical concept) can occur if the quantum state is a solution of the minimum uncertainty relation \cite{Schiff}, called the minimum uncertainty state. Because there was a transition from the quantum to the classical states in the early universe, it does not rule out the possibility that the initial state was a state with measurement accuracy closest to classical physics, namely, a state of minimal uncertainty. Observables that can be analyzed from gravitational waves are needed to obtain an initial state of minimum uncertainty of primordial gravitons. In this research, the observable that will be used is observables that describe the polarization intensity of gravitational waves with measurement operators that can be represented in the form of the Stokes operator \cite{Stokes}. There are four Stokes operators, each of which will define the intensity of linear, circular, and total polarization of the gravitons. By using these operators, the minimum uncertainty state should be a unique quantum state, for example, entanglement state as in the research \cite{Nha, Aragone, Hillery, Brif}. The entanglement that would emerge from the minimal uncertainty based on Stokes operators should be the entanglement between the polarization modes of the primordial gravitational waves. The existence of a non-classical correlation between these polarization modes will not appear if the initial state is a Bunch-Davies vacuum. Even though, in the radiation domination era, the Bunch-Davies vacuum can be seen as an entanglement, that entanglement is between the wave number modes $\mathbf{k}$ and $-\mathbf{k}$, not between the polarization modes of the primordial gravitational waves. This research will show how the initial state with non-classical correlation between the polarization modes impacts the decoherence due to the interaction of primordial gravitons with the interferometer mirrors. But before that, to prove that the initial state with minimal uncertainty conditions has that non-classical correlation, we will calculate a quantity called quantum discord \cite{Horodecki, Zurek}. Several similar studies have also been carried out to examine the properties of primordial gravitons if these particles have non-classical correlations \cite{Kanno5, Kanno6}. However, in those studies, the quantum correlation was a non-classical correlation between graviton particles and photons caused by interaction with the electromagnetic field. 

The organization of this paper is as follows. First, in section \ref{sec2}, we will explain the possible initial states of minimum uncertainty based on Stokes operator. Next, in section \ref{sec3}, the quantum discord quantity will be calculated to show that the initial minimum uncertainty states of primordial graviton will have a quantum correlation between the polarization modes. In section \ref{sec4}, before studying the decoherence of the interferometer mirrors due to primordial gravitons with an initial state of minimum uncertainty in section \ref{sec5}, we will first review the quantum noise operator of the graviton, which will be used to calculate the decoherence. Then, in section \ref{sec6} is the conclusion. This paper will include an Appendix, which contains details of the additional calculations needed to calculate quantum discord in section \ref{sec3} in part \ref{ApenA}, as well as in part \ref{ApenB} containing the influence functional method used to determine the decoherence functional due to the noise of gravitons, which will be used in the section \ref{sec4} to calculate the decoherence time.

\section{Minimum Uncertainty of Graviton}\label{sec2}
In quantum mechanics, if there are two hermitian operators $\hat{X}_1$ and $\hat{X}_2$, the Heisenberg uncertainty relation can be expressed as
$\big(\Delta\hat{X}_1\big)^2 \big(\Delta\hat{X}_2\big)^2\geq\frac{1}{4}|\braket{\big[\hat{X}_1,\hat{X}_2\big]} |^2$.
A quantum state can be said to fulfil its minimal conditions or minimal uncertainty if the quantum state is a solution to the eigenequation \cite{Schiff}
\begin{equation}\label{eq:MinUncer}
\big(\hat{X}_1+i\lambda\hat{X}_2\big)\ket{\psi}=\beta\ket{\psi},
\end{equation}
where $\beta$ is the eigenvalue and $\lambda$ is the squeezing parameter of the solution of the eigenstate. The squeezed parameter $\lambda$ will determine the initial quantum state of the system in its minimal uncertainty condition. To see why $\lambda$ is defined as the squeezing parameter, it can be explained by looking for the variance of the $\hat{X}_1$ and $\hat{X}_2$ operators based on the minimal uncertainty relation (\ref{eq:MinUncer}). The variance can be written as follows
\begin{equation}\label{eq:Varians1}
\big(\Delta \hat{X}_1\big)^2=\frac{1}{2}\lambda\,\,\,|\bra{\psi}[\hat{X}_1,\hat{X}_2]\ket{\psi}|
\end{equation} 
and
\begin{equation}\label{eq:Varians2}
(\Delta \hat{X}_2)^2=\frac{1}{2\lambda}\,\,\,\big|\bra{\psi}[\hat{X}_1,\hat{X}_2]\ket{\psi}\big|.
\end{equation}
It can be seen that when $\lambda<1$, the variance of equation (\ref{eq:Varians2}) will be greater than equation (\ref{eq:Varians1}). Based on the definition of a squeezed state \cite{Knight, Fox}, a quantum state can be said to be squeezed if the variance of one operator that satisfies the minimum uncertainty is smaller than the variance of the other operators, which means that when $\lambda<1$ the quantum state will be squeezed. It also applies when $\lambda>1$ except that the variance of equation (\ref{eq:Varians1}) will be larger than equation (\ref{eq:Varians2}). When $\lambda=0$, the stated will not be squeezed because the two variances are the same. Furthermore, for simplicity of calculation, the value of $\lambda$ will be limited to $0\leq\lambda\leq1$.

To find out the minimum uncertainty relation of the primordial graviton, the arbitrary operators $\hat{X}_1$ and $\hat{X}_2$ in equation (\ref{eq:MinUncer}) are chosen to be operators that define the polarization of gravitational waves. Like the photon, the polarization of gravitational waves can be defined by the Stokes operators. Using the representation of annihilation and creation operators of graviton $(\hat{a}^{A}_{\mathbf{k}},\,\,\hat{a}^{A\dagger}_{\mathbf{ k}})$, where this creation annihilation operator will satisfy $\hat{a}^{A}_{\mathbf{k}}\ket{0^A}=0$, the commutation relation $ [\hat{a}^{A}_{\mathbf{k}},\,\,\hat{a}^{A'\dagger}_{\mathbf{l}}]=\delta^{A\,A'} \delta_{\mathbf{k}\,\mathbf{l}}$ and the index $A$ describe the polarization mode of the graviton which can be $\otimes$ and $\oplus$. Thus, the Stokes operators can be expressed as
\begin{eqnarray}
\hat{\mathcal{S}}^{(0)}_k&=&\hat{a}^{\oplus\dagger}_{\mathbf{k}}\hat{a}^{\oplus}_{\mathbf{k}}+\hat{a}^{\otimes\dagger}_{\mathbf{k}}\hat{a}^{\otimes}_{\mathbf{k}},\,\,\,\,\,\,\,\hat{\mathcal{S}}^{(1)}_k=\hat{a}^{\oplus\dagger}_{\mathbf{k}}\hat{a}^{\oplus}_{\mathbf{k}}-\hat{a}^{\otimes\dagger}_{\mathbf{k}}\hat{a}^{\otimes}_{\mathbf{k}}\nonumber\\
\hat{\mathcal{S}}^{(2)}_k&=&\hat{a}^{\oplus\dagger}_{\mathbf{k}}\hat{a}^{\otimes}_{\mathbf{k}}+\hat{a}^{\oplus}_{\mathbf{k}}\hat{a}^{\otimes\dagger}_{\mathbf{k}},\,\,\,\,\,\,\,\hat{\mathcal{S}}^{(3)}_k=i\big(\hat{a}^{\oplus}_{\mathbf{k}}\hat{a}^{\otimes\dagger}_{\mathbf{k}}-\hat{a}^{\oplus\dagger}_{\mathbf{k}}\hat{a}^{\otimes}_{\mathbf{k}}\big).
\end{eqnarray}
The operator $\hat{\mathcal{S}}^{(0)}_k$ defines the total intensity of all modes of the gravitons. $\hat{\mathcal{S}}^{(1)}_k$ and $\hat{\mathcal{S}}^{(2)}_k$ are intensity operators of linear polarization. Meanwhile, $\hat{\mathcal{S}}^{(3)}_k$ is an operator for measuring the intensity of circular polarization. The last three Stokes operator will have a commutation relation $[\hat{\mathcal{S}}^{(i)}_k,\hat{\mathcal{S}}^{(j)}_k]=2i\epsilon^{ijk}\hat{\mathcal{S}}^{(k)}_k$ which satisfies the SU(2) algebra with $(i,j,k)\equiv 1,2,3$. However, for operator $\hat{\mathcal{S}}^{(0)}_k$, the commutation relations with other operators will always be zero, which means the operator $\hat{\mathcal{S}}^{(0)}_k$ cannot be used to study minimal uncertainty.

The minimum uncertainty states from these operators will have a meaning as quantum states that minimize the uncertainty of measuring the intensity of gravitational waves with different types of polarization based on the Heisenberg uncertainty principle. Apart from being a squeezed parameter, in this case, the parameter $\lambda$ would also be a degree of freedom that determines the deviation between the variances of gravitational wave intensity measurements with different types of polarization.

Furthermore, by choosing arbitrary operators $\hat{X}_1$ and $\hat{X}_2$ as the Stokes operators, then the minimum uncertainty states of the primordial graviton, which will be used as initial states of the early universe, can be obtained. First we chose the arbitrary operators $\hat{X}_1=\hat{S}^{(2)}_k$ and $\hat{X}_2=\hat{S}^{(3)}_k$ . Substitute into equation (\ref{eq:MinUncer}), the minimum uncertainty relation will be obtained as
\begin{equation}\label{eq:Eigen}
\bigg[(1+\lambda)\hat{a}^{\oplus\dagger}_{\mathbf{k}}\hat{a}^{\otimes}_{\mathbf{k}}+(1-\lambda)\hat{a}^{\oplus}_{\mathbf{k}}\hat{a}^{\otimes\dagger}_{\mathbf{k}}\bigg]\ket{\psi}=\beta\ket{\psi}.
\end{equation}
As explained in the first pharagraph, there are three possible values of $\lambda$ which are $\lambda=0$,   $\lambda=1$ and $0<\lambda<1$. For $\lambda=0$ the minimum uncertainty relation become
\begin{equation}\label{eq:Eigen0}
\bigg[\hat{a}^{\oplus\dagger}_{\mathbf{k}}\hat{a}^{\otimes}_{\mathbf{k}}+\hat{a}^{\oplus}_{\mathbf{k}}\hat{a}^{\otimes\dagger}_{\mathbf{k}}\bigg]\ket{\psi}=\beta\ket{\psi}.
\end{equation}
To obtain the solution, the general form of the quantum state is taken $\ket{\psi}=\sum_{n,m}\mathcal{C}_{nm}\ket{n^{\oplus}_{\mathbf{k}},m^{\otimes}_{\mathbf{k}}}$. Where $\mathcal{C}_{nm}$ is a constant that depends on the value of $n$ and $m$. Substitute back into equation (\ref{eq:Eigen0}), then multiply by $\bra{i^{\oplus}_{\mathbf{k}},j^{\otimes}_{\mathbf{k}}}$ from the left, we got
\begin{align}
\sqrt{i(j+1)}\mathcal{C}_{(i-1)\,(j+1)}+\sqrt{(i+1)(j-1)}\,\,\mathcal{C}_{(i+1)\,(j-1)}=\beta\,\,\mathcal{C}_{ij}.
\end{align}
The constant $\beta$ is chosen to be the maximum number of gravitons in the initial state  $(N=n+m)$, then the constant $\mathcal{C}_{nm}$ can be found to be $\begin{pmatrix}
N \\
n
\end{pmatrix}^{1/2}$. So, the solution to the eigenstate become
\begin{eqnarray}\label{eq:Entang1}
\ket{\psi}&=&\sum_{n=0}^{N}\,\,\frac{1}{2^{N/2}}\,\,\begin{pmatrix}
N \\
n
\end{pmatrix}^{1/2}\ket{(N-n)^{\oplus}_{\mathbf{k}},\,\,n^{\otimes}_{\mathbf{k}}}.
\end{eqnarray}
where $1/2^{N/2}$ appears due to normalization. This state is an entanglement between the polarization of modes $\oplus$ and $\otimes$. As explained in the introduction, the state of minimal uncertainty from the polarization of gravitational waves can be a state of entanglement. It is well known that the entanglement state is a unique condition in quantum physics that does not exist in classical physics. This state should have a non-classical correlation between the two polarization modes.

For $\lambda=1$, the equation (\ref{eq:Eigen}) becomes $2\hat{a}^{\oplus\dagger}_{\mathbf{k}}\hat{a}^{\otimes}_{\mathbf{k}}\ket{\psi}=\beta\ket{\psi}$. There will be a solution if $\beta=0$ with the eigenstate being a vacuum for the polarization of mode $\otimes$ and can be any state for the mode $\oplus$ ($\ket{\psi}=\ket{\phi^{\oplus},\,0^{\otimes}}$ or for simplicity is chosen $\ket{\psi}=\ket{0^{\oplus},0^{\otimes}}$). Likewise if the $\lambda$ constant is $0<\lambda<1$, the solution will also only be a vacuum state ($\ket{\psi}=\ket{0^{\oplus},0^{\otimes}}$) with $\beta=0$. So, the vacuum state will be the solution for $0\leq\lambda<1$.

To obtain a general solution of the equation (\ref{eq:Eigen}), each solution for each value of $\lambda$ will be expressed in the form of a linear combination or superposition. In this case, there are only two solutions, which are a vacuum state and the entanglement (\ref{eq:Entang1}). So the general solution of the minimum uncertainty relation (\ref{eq:Eigen}) will be a superposition between a vacuum state and entanglement (\ref{eq:Entang1}), where mathematically, it can be written as follows
\begin{equation}\label{eq:2.9}
\ket{\psi}=f_1(\lambda)\,\,\ket{0^{\oplus}, 0^{\otimes}}+f_2(\lambda)\sum_{n=0}^{N}\,\,\frac{1}{2^{N/2}}\,\,\begin{pmatrix}
N \\
n
\end{pmatrix}^{1/2}\ket{(N-n)^{\oplus}_{\mathbf{k}},\,\,n^{\otimes}_{\mathbf{k}}},
\end{equation}
with $f_1(\lambda)$ and $f_2(\lambda)$ are functions that depend on $\lambda$. Because the equation (\ref{eq:2.9}) must be normalized $(\braket{\psi|\psi}=1)$, then the two functions will satisfy the relation $|f_1(\lambda)|^2+|f_2( \lambda)|^2=1$. Based on this, it is chosen
\begin{equation}
f_2(\lambda)=\Big(1-|f_1(\lambda)|\Big)^{1/2}.
\end{equation}
When $\lambda=1$ the equation (\ref{eq:2.9}) have to be the equation (\ref{eq:Entang1}), it means the function $f_1(\lambda)=0$ and $f_2(\lambda)=1$. On the other hand, when $0\leq\lambda<1$, the equation (\ref{eq:2.9}) must be a vacuum state, meaning that $f_1(\lambda)=1$ and $f_2(\lambda)=0$. So, it can be said that $f_1(\lambda)$ is a function that resembles a ladder function with the highest value being 1, and if $\lambda\simeq1$ then $f_1(\lambda)\simeq0$. For this reason, the function $f_1(\lambda)$ will be chosen to be a hyperbolic tangent function so that the equation (\ref{eq:2.9}) becomes
\begin{eqnarray}\label{eq:superposition1}
\ket{\psi}=\tanh u\,\,\ket{0^{\oplus},0^{\otimes}}+(1-\tanh^2 u)^{1/2}\sum_{n=0}^{N}\,\,\frac{1}{2^{N/2}}\,\,\begin{pmatrix}
N \\
n
\end{pmatrix}^{1/2}\ket{(N-n)^{\oplus}_{\mathbf{k}},\,\,n^{\otimes}_{\mathbf{k}}}.
\end{eqnarray}
Where $u\equiv c\,\lambda$ and $c$ is an arbitrary positive real number with value $c>>1$. According to this definition, the constant $u$ has a range of values of $0\leq u < \infty$. This general solution should also have non-classical correlations between the polarization modes. Interestingly, if we consider the density matrix of this state for one of the polarization modes, then the density matrix will have non-diagonal elements. This contrasts with the entanglement solution in equation (\ref{eq:Entang1}), which only has diagonal elements if the density matrix is considered for one of the polarization modes. As shown in the equations (\ref{eq:3.8}) and (\ref{eq:Dens1}). The solution (\ref{eq:superposition1}) will also be satisfied if $\hat{X}_1=\hat{\mathcal{S}}^{(2)}_k$ and $\hat{X}_2=\hat{\mathcal{S}}^{(1)}_k$.

Not only the entanglement (\ref{eq:Entang1}) but there is another entanglement state that also satisfies the minimum uncertainty relation. Selected $\hat{X}_1=\hat{\mathcal{S}}^{(1)}_k$ and $\hat{X}_2=\hat{\mathcal{S}}^{(3)} _k$, substitute to the minimum uncertainty equation  (\ref{eq:Eigen}), then
\begin{equation}\label{eq:Eigen1}
\bigg[\hat{a}^{\oplus\dagger}_{\mathbf{k}}\hat{a}^{\oplus}_{\mathbf{k}}-\hat{a}^{\otimes\dagger}_{\mathbf{k}}\hat{a}^{\otimes}_{\mathbf{k}}-\lambda\big(\hat{a}^{\oplus}_{\mathbf{k}}\hat{a}^{\otimes\dagger}_{\mathbf{k}}-\hat{a}^{\oplus\dagger}_{\mathbf{k}}\hat{a}^{\otimes}_{\mathbf{k}}\big)\bigg]\ket{\psi}=\beta\ket{\psi}.
\end{equation}
The solution of this equation can be found using the same method as before. Chose $\lambda=1$ and using the general solution $\ket{\psi}=\sum_{n,m}\mathcal{C}_{nm}\ket{n_{\mathbf{k}}^{\oplus},m_{\mathbf{k}}^{\otimes}}$, then substituted this solution and multiplied $\bra{i^{\oplus}_{\mathbf{k}},j^{\otimes}_{\mathbf{k}}}$ from the left, we got
\begin{equation}
\sqrt{i(j+1)}\,\,\mathcal{C}_{(i-1)(j+1)}-\sqrt{j(i+1)}\,\,\mathcal{C}_{(i+1)(j-1)}=\Big(\beta-(i-j)\Big)\,\,\mathcal{C}_{ij},
\end{equation}
the expression for the constant $\mathcal{C}_{nm}$ can be found if the value $\beta=0$. Where is the solution to the constant 
 $\mathcal{C}_{nm}$ is $(-1)^n\begin{pmatrix}
N\\
n
\end{pmatrix}^{1/2}$, so
\begin{equation}\label{eq:Entang2}
\ket{\psi}=\sum_{n=0}^{N}\,\,\frac{1}{2^{N/2}}\,\,(-1)^n\begin{pmatrix}
N \\
n
\end{pmatrix}^{1/2}\ket{(N-n)^{\oplus}_{\textbf{k}},\,\,n^{\otimes}_{\textbf{k}}},
\end{equation}
with $1/2^{N/2}$ appears due to normalization. This solution is similar to the entanglement (\ref{eq:Entang1}), only there is a factor $(-1)^n$ in the solution. For $0\leq\lambda<1$ the solution is a vacuum. So, the general solution of the equation (\ref{eq:Eigen1}), using the superposition expression, can be written as follows
\begin{eqnarray}\label{eq:superposition2}
\ket{\psi}&=&\tanh u\,\,\ket{0^{\oplus},0^{\otimes}}+(1-\tanh^2 u)^{1/2}\sum_{n=0}^{N}\frac{1}{2^{N/2}}\,\,(-1)^n\,\,\begin{pmatrix}
N \\
n
\end{pmatrix}^{1/2}\ket{(N-n)^{\oplus}_{\textbf{k}},\,\,n^{\otimes}_{\textbf{k}}},
\end{eqnarray}
where the constant $u$ will be defined differently, namely $u\equiv c\,(1-\lambda)$ which has the range $0\leq u<\infty$. This solution would also be satisfied when $\hat{X}_1=\hat{\mathcal{S}}^{(1)}_k$ and $\hat{X}_2=\hat{\mathcal{S}}^{(3)}_k$. The following section will show that the minimum uncertainty initial states will have a non-classical correlation between the polarization modes. This will be proven based on quantum discord calculations.

\section{Quantum Discord of The Minimum Uncertainty Initial State}\label{sec3}
Quantum discord will be calculated to show that the initial state of equation (\ref{eq:superposition1}) has a quantum (non-classical) correlation between the polarization of modes $\otimes$ and $\oplus$. Quantum discord is a quantity used to measure the non-classical correlation  of two or more systems. Usually, non-classical correlations are measured using entanglement measurements \cite{Horodecki}, but in the case of mixed bipartite entanglement, the measurements cannot characterize non-classical correlations well. Therefore, in 2001 H. Oliver and W.H. Zurek developed a non-classical correlation measurement called quantum discord \cite{Zurek}.

Quantum discord is defined as the maximum difference of quantum mutual information with and without von Neumann projection measurements applied to one part of the system. To begin the discussion we will consider a bipartite case, where there are quantum systems $A$ and $B$ with the total density matrix can be expressed as $\rho^{AB}$. Quantum mutual information can be calculated by adding the entropy of systems $A$ ($S_A(\rho^A)$) and $B$ ($S_B(\rho^B)$) and then subtracting the entropy for the total system ($S_{AB}(\rho^{AB})$). Because the system under consideration is a quantum system, the entropy used is von Neumann entropy $\big(S_x(\rho^x)=-Tr(\rho^x\,\,\log\,\,\rho^x)\big)$. So the mutual information of two systems A and B can be written as follows
\begin{equation}\label{eq:3.1}
\mathcal{I}(\rho^{AB})=S_A(\rho^A)+S_B(\rho^B)-S_{AB}(\rho^{AB}).
\end{equation}
This mutual information is calculated without using von Neumann projective measuement. Meanwhile, if von Neuman projective measurement is used (for system A), then the mutual information is
\begin{equation}\label{eq:3.2}
\mathcal{J}(\rho^{AB})=S_B(\rho^B)-\text{min}\Big\{S_{B|\Pi^A}(\rho^{AB})\Big\}
\end{equation}
The mutual information of the equation (\ref{eq:3.1}) could be different from the equation (\ref{eq:3.2}). The difference arises due to the second term of the equation (\ref{eq:3.2}). This term is referred to as the conditional entropy. Where conditional entropy is the entropy produced by one part of the system (in this case system $B$) with the other part (system $A$) having a certain value. To have a certain value for system $A$, a von Neumann projection operator ($\Pi^A_j$) is used, which satisfies the positive operator valued measure (POVM), namely $\sum_j\,\Pi^A_j=1$. If we explain further, the conditional entropy in the equation (\ref{eq:3.2}) can be written as
\begin{equation}
S_{B|\Pi^A}(\rho^{AB})=\sum_{j}P_j \,\,S(\rho^B_j)
\end{equation}
where $P_j$ is the probability for each $j$ or $P_j=Tr\big\{(\Pi^A_j\otimes I)\rho^{AB}(\Pi^A_j\otimes I)\big\}$ and the density matrix $\rho^B_j$ is
\begin{equation}
\rho^B_j=\frac{1}{P_j}\,\,Tr_A\Big\{(\Pi^A_j\otimes I)\rho^{AB}(\Pi^A_j\otimes I)\Big\}.
\end{equation}
The minimum index in the second term of the equation (\ref{eq:3.2}) refers to the choice of von Neumann projection operator. Where the form of the projection operator is chosen to produce the lowest conditional entropy so that the mutual information equation (\ref{eq:3.2}) will be greater.

The quantum discord $\big(\mathcal{D}(\rho^{AB})\big)$ can be defined as the difference of the equations (\ref{eq:3.1}) and (\ref{eq:3.2}). Mathematically quantum discord can be written as follows
\begin{eqnarray}\label{eq:3.5}
\mathcal{D}(\rho^{AB})&=&\mathcal{I}(\rho^{AB})-\mathcal{J}(\rho^{AB})\nonumber\\
&=&S_A(\rho^A)-S_{AB}(\rho^{AB})+\text{min}\Big\{S_{B|\Pi^A}(\rho^{AB})\Big\}
\end{eqnarray}
This quantity will always be worth $\geq\,0$. If the quantum discord is more than zero (mutual information (\ref{eq:3.1}) is different from (\ref{eq:3.2})) it can be said that there is a quantum correlation between systems $A$ and $B$. Measurement of the von Neumann projection on system $A$ will change the state of system $B$ if there is a quantum correlation between the two systems so that the difference between the two mutual information calculations will be more than zero. If the quantum discord quantity is zero, then it can be said that there is no quantum correlation between systems $A$ and $B$ or the correlation is classical.

\begin{figure*}
	\centering 
	\includegraphics[width=11cm]{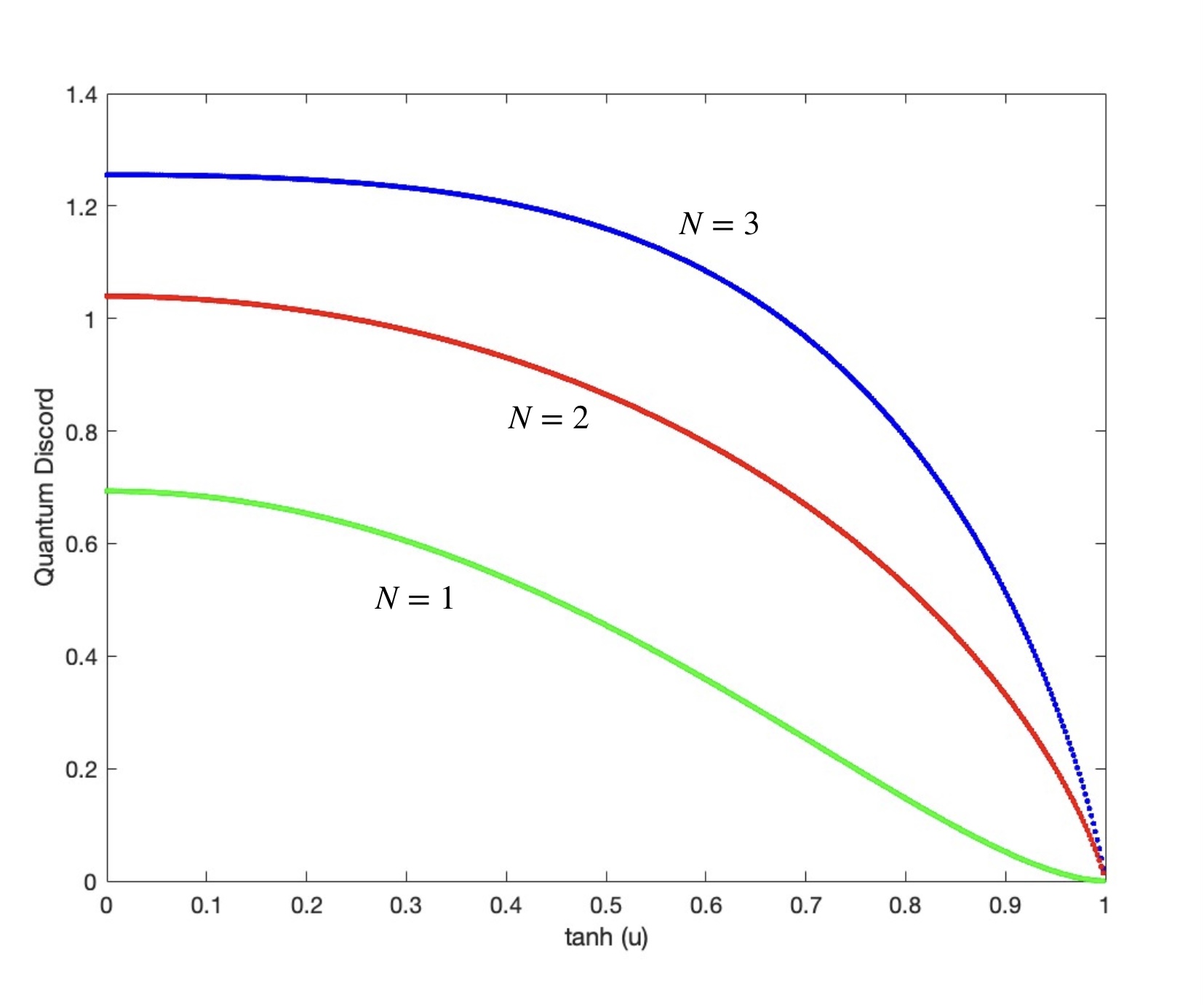}
	\caption{\label{fig1} The quantum discord quantity from the initial state of equation (\ref{eq:superposition1}) with the maximum number of initial gravitons $N=1, 2,$ and $3$. }
\end{figure*}

If the quantum discord equation (\ref{eq:3.5}) is expressed for a bipartite density matrix between the polarization modes $\otimes$ and $\oplus$ $(\rho^{\oplus\otimes})$, then
\begin{equation}
\mathcal{D}(\rho^{\oplus\otimes})=S_{\oplus}(\rho^{\oplus})-S_{\oplus\otimes}(\rho^{\oplus\otimes})+\text{min}\Big\{S_{\otimes|\Pi^{\oplus}}(\rho^{\oplus\otimes})\Big\}.
\end{equation}
Recall the initial state of equation (\ref{eq:superposition1}) and (\ref{eq:superposition2}). The density matrix of this state can be written
\begin{eqnarray}
\rho^{\oplus\otimes}_{\pm}&=&\ket{\psi}\bra{\psi}\nonumber\\
&=&\tanh^2 u\ket{0^{\oplus}, 0^{\otimes}}\bra{0^{\oplus}, 0^{\otimes}}+\big(1-\tanh^2 u\big)\sum^N_{m,n=0}\frac{1}{2^N}(\pm1)^{n+m}
\begin{pmatrix}
N\\
n
\end{pmatrix}^{1/2}\,\,
\begin{pmatrix}
N\\
m
\end{pmatrix}^{1/2}\,\,\nonumber\\
&\,\,&\times\ket{(N-n)^{\oplus}_{\mathbf{k}},n^{\otimes}_{\mathbf{k}}}\bra{(N-m)^{\oplus}_{\mathbf{k}},m^{\otimes}_{\mathbf{k}}}+\tanh u\,\big(1-\tanh^2 u\big)^{1/2}\,\,\frac{1}{2^{N/2}}\sum^N_{n,m=0}\nonumber\\
&\,\,&\times\bigg[(\pm1)^n\,\begin{pmatrix}
N\\
n
\end{pmatrix}^{1/2}\ket{(N-n)^{\oplus}_{\mathbf{k}},n^{\otimes}_{\mathbf{k}}}\bra{0^{\oplus}, 0^{\otimes}}+(\pm1)^m\,\begin{pmatrix}
N\\
m
\end{pmatrix}^{1/2}\ket{0^{\oplus}, 0^{\otimes}}\bra{(N-m)^{\oplus}_{\mathbf{k}},m^{\otimes}_{\mathbf{k}}}\bigg],\nonumber\\
\end{eqnarray}
where the index $"+"$ represents density matrix for equation (\ref{eq:superposition1}) and $"-"$ represents equation (\ref{eq:superposition2}). The density matrix in mode $\oplus$ for both equation become
\begin{eqnarray}\label{eq:3.8}
\rho^{\oplus}&=&Tr_{\otimes}\big(\ket{\psi}\bra{\psi}\big)\nonumber\\
&=&\tanh^2 u \ket{0^{\oplus}}{\bra{0^{\oplus}}}+(1-\tanh^2 u)\,\,\sum_{n=0}^{N}\,\,\frac{1}{2^{N}}\,\,\begin{pmatrix}
N \\
n
\end{pmatrix}\ket{(N-n)^{\oplus}_{\mathbf{k}}}\bra{(N-n)^{\oplus}_{\mathbf{k}}}\nonumber\\
&\,\,&+\frac{1}{2^{N/2}}\tanh u\,\,(1-\tanh^2 u)^{1/2}\,\,\Big[\ket{0^{\oplus}}{\bra{N^{\oplus}_{\mathbf{k}}}}+\ket{N^{\oplus}_{\mathbf{k}}}{\bra{0^{\oplus}}}\Big]\
\end{eqnarray}
Using this density matrix, the von Neumann entropy $S_{\oplus\otimes}(\rho^{\oplus\otimes})$ will be zero because $\rho^{\oplus\otimes}_{\pm}$ is a pure state. Likewise with the entropy $S_{\otimes|\Pi^{\oplus}}(\rho^{\oplus\otimes})$, as shown in the Appendix (\ref{ApenA}). Where if $\rho^ {\oplus\otimes}_{\pm}$ is a pure state, so $\rho^{\otimes}_j$ will also be a pure state, which means that the quantum discord quantity will only depend on the entropy $S_{\oplus}(\rho^{\oplus})$. The amount of entropy can be obtained by calculating the eigenvalues of the density matrix (\ref{eq:3.8}) $\big(S_{\oplus}(\rho^{\oplus})=-\sum^N_{j=0}\eta_j\,\ln\eta_j$, where $\eta_j$ is the eigenvalue$\big)$. For the maximum number of initial gravitons $N=1,2$ and $3$ , the results obtained are as shown in figure (\ref{fig1}). It can be seen that the initial state of equation (\ref{eq:superposition1}) will have a quantum discord greater than zero if $\tanh u\neq1$, when $\tanh u=1$, the initial state will be a Bunch-Davies vacuum, where this state has no quantum correlation between the mode $\oplus$ and $\otimes$. Apart from that, it can also be seen that the more total gravitons (N), the greater the quantum nature of the correlation between $\oplus$ and $\otimes$ modes.

\section{Quantum Noise from Graviton}\label{sec4}
This section will briefly review quantum noise caused by gravitons based on research from  S. Kanno \textit{et al.} \cite{Kanno3}. Quantum noise is a consequence that arises due to an interaction between the system and the environment within the framework of quantum mechanics. His research assumes that the system is an interferometer with two mirrors in the gravitational wave environment. If the gravitational waves are quantized (in the form of graviton particles), it is expected that there will be interactions between the system and the environment. So that the quantum noise would appear when the interferometer makes the measurements. 

The quantum noise operator can be mathematically obtained by assuming the interferometer mirror is a particle in the Fermi normal coordinates system. In this coordinates system, there are two particles, where one of those (a particle with a timelike geodesic $\gamma_{\tau}$) will be used as a frame of reference. While the other particles (particles with timelike geodesic $\gamma_{\tau'}$) are test particles that have a position $x^i(t)=\xi^i(t)$ to the reference frame at point $P(0,t )$. The Fermi normal coordinates are used because of the Einstein equivalence principle, where if only one particle is considered, the effect of gravitational waves cannot be affected.

Next, consider the geodesic action of the test particles. By using the metric in Fermi coordinates for the second order of $x^i$ as
\begin{eqnarray}
ds^2&\simeq&(-1-R_{0i0j}x^ix^j)dt^2-\frac{4}{3}R_{0jik}x^jx^kdtdx^i+\Big(\delta_{ij}-\frac{1}{3}R_{ikjl}x^kx^l\Big)dx^idx^j.
\end{eqnarray}
The test particle action will be obtained as
\begin{equation}
S_p\simeq\int_{\gamma_{\tau'}}dt\Big[\frac{m}{2}\dot{{\xi^i}^2}-\frac{m}{2}R_{0i0j}(0,t)\xi^i\xi^j\Big].
\end{equation}
Because the mirror particle is in an environment with gravitational waves, the Riemann tensor $R_{0i0j}$ can be obtained by using a flat metric space and time with small perturbations $h_{ij}$ that satisfies the transverse traceless gauge $(\partial_ih^{ij}= 0$, and $h^i_i=0 )$. So if the perturbation $h_{ij}$ is represented in Fourier space, we get action
\begin{equation}
S_p\simeq\int_{\gamma_{\tau'}}dt\bigg[\frac{m}{2}\dot{{\xi^i}^2}-\frac{m}{2}\frac{\kappa}{\sqrt{V}}\sum_{A}\sum_{\textbf{k}\leq\Omega_m}\ddot{h}^A_{\textbf{k}}(t)\,\,e^A_{\textbf{k},ij}\,\,\xi^i\xi^j\bigg],
\end{equation}
where $\sum_{\textbf{k}\leq\Omega_m}$ represents the sum of $\textbf{k}$ modes, with UV cutoof $\Omega_m\sim\xi^{-1}$. In general, the total action can be written as the sum of the actions of the test particles plus the gravitational action $(S=S_p+S_g)$, which can be written as follows
\begin{eqnarray}\label{eq:TotAction}
S&\simeq&\int dt\sum_{\textbf{k},A}\bigg[\frac{1}{2}\dot{h}^A_{\textbf{k}}(t)\dot{h}^{*A}_{\textbf{k}}(t)-\frac{1}{2}k^2h^A_{\textbf{k}}(t)h^{*A}_{\textbf{k}}(t)\bigg]+\int dt\bigg[\frac{m}{2}\dot{{\xi^i}^2}-\frac{m}{2}\frac{\kappa}{\sqrt{V}}\sum_{A}\sum_{\textbf{k}\leq\Omega_m}\ddot{h}^A_{\textbf{k}}(t)\,\,e^A_{\textbf{k},ij}\,\,\xi^i\xi^j\bigg].\nonumber\\
\end{eqnarray}
The gravitational action in the first term is just an action with the equation of motion similar to a harmonic oscillator. By using the interaction picture, the field $h^A_{I,\textbf{k}}(t)$ can be quantized by defining it in terms of the annihilation and creation operators
\begin{equation}
\hat{h}^A_{\textbf{k},I}(t)=\hat{a}^A_{\textbf{k}}v_k(t)+\hat{a}^{A\dagger}_{-\textbf{k}}v^{*}_k(t)
\end{equation}
Where $v_k(t)$ is the solution to the equation of motion which has the form $v_k(t)=(1/\sqrt{2k})e^{-ikt}$ and the annihilation and creation operators would satisfy $[\hat{a}^A_{\textbf{k}}\hat{a}^{A'\dagger}_{\textbf{k}'}]=\delta_{A,A'}\delta_{\textbf{k}\textbf{k}'}$. Next, the field $\hat{h}^A_{I,\textbf{k}}(t)$ will be divided into two parts, namely the "classic" part $\Big(h_{cl,\textbf{k}}(t )\equiv \braket{\hat{h}^A_{\textbf{k},I}(t)}\Big)$ and the "quantum" part which can be written as
\begin{equation}
\delta\hat{h}^A_{\textbf{k},I}(t)=\hat{h}^A_{\textbf{k},I}(t)-h_{cl,\textbf{k}}(t).
\end{equation}
From this equation, one could say that $\delta\hat{h}^A_{I,\textbf{k}}(t)$ will define the quantum fluctuations of gravitational waves due to graviton particles. To review the expression of the quantum noise, we will look for the equation of motion of the test particle from the action of equation (\ref{eq:TotAction}). Then we will look for terms from the equation of motion where the quantum fluctuations $\delta\hat{h}^A_{I,\textbf{k}}(t)$ will have an effect. From this term, we can define quantum noise caused by graviton particles. The equation of motion of the test particle where $\xi^i$ is promoted as operator $\hat{\xi^i}$ can be written

\begin{align}\label{eq:Motion}
\ddot{\hat{\xi}}^i(t)&-\frac{1}{2}\ddot{h}^{cl}_{ij}(0,t)\hat{\xi}^j(t)+\frac{\kappa^2m}{40\pi}\bigg[\delta_{ik}\delta_{jl}+\delta_{il}\delta_{jk}-\frac{2}{3}\delta_{ij}\delta_{kl}\bigg]\hat{\xi}^j(t)\frac{d^5}{dt^5}\Biggl\{\hat{\xi}^k(t)\hat{\xi}^l(t)\Biggr\}\nonumber\\
&=-\frac{\kappa}{\sqrt{V}}\sum_{A}\sum_{\textbf{k}\leq\Omega_m}k^2\,\,e^A_{\textbf{k},ij}\delta\hat{h}^A_{\textbf{k},I}(t)\,\,\xi^j+\frac{\kappa^2m}{20\pi^2}\Omega_m\bigg[\delta_{ik}\delta_{jl}+\delta_{il}\delta_{jk}-\frac{2}{3}\delta_{ij}\delta_{kl}\bigg]\hat{\xi}^j(t)\frac{d^4}{dt^4}\Biggl\{\hat{\xi}^k(t)\hat{\xi}^l(t)\Biggr\}
\end{align}

This result is similar to the calculation obtained by Parikh \textit{et. al} \cite{PWZ}. The term where quantum fluctuations will affect is in the first term on the right side of the equation (\ref{eq:Motion}), so quantum noise can be defined as follows
\begin{equation} \label{Noise}
\hat{N}_{ij}(t)=\frac{\kappa}{\sqrt{V}}\sum_{A}\sum_{\textbf{k}\leq\Omega_m}k^2\,\,e^A_{\textbf{k},ij}\delta\hat{h}^A_{\textbf{k},I}(t).
\end{equation}
Where the quantum noise caused by graviton $ \hat{N}_{ij}(t)$ will always be there as long as gravitational waves are quantized.

\section{Decoherence Induced by Minimum Uncertainty Initial State of Primordial Graviton}\label{sec5}

To study decoherence due to the noise of primordial gravitons which have a minimum uncertainty initial state, the experimental setup of the system is chosen to have a form similar to research \cite{Kanno4}. In that experimental setup, the Michelson equal arm interferometer, which has two macroscopic suspended mirrors at the end of each interferometer arm, is used. When the laser interferometer is fired, there will be two possible paths towards mirror 1 or mirror 2. Furthermore, when a photon from the laser interferometer hits one of the mirrors, it is assumed that an oscillation will occur, and it is described as the following semiclassical state.
\begin{eqnarray}\label{eq:Setup}
\vec{\xi}_1(t)&=&(\xi_1,0,0),\,\,\,\,\,\,\,\,\,\,\,\,\,\,\xi_1=A\cos \omega t\nonumber\\
\vec{\xi}_2(t)&=&(0,\xi_1,0),\,\,\,\,\,\,\,\,\,\,\,\,\,\,\xi_2=A\cos \omega t.
\end{eqnarray}
Where $\omega$ and $A$ are the frequency and amplitude that can be generated from the oscillations of each mirror. In the Hilbert space representation, there will be two possible bases, namely $\ket{0}$ which is the basis vector when the photon does not hit the mirror, and $\ket{\vec{\xi_i}}$ is the basis vector when the photon hits one of the mirrors $(i=1,2)$. Overall, the system's quantum state can be expressed in the superposition form as follows
\begin{equation}
\ket{\Psi(t_i)}_m=\frac{1}{\sqrt{2}}\bigg(\ket{\vec{\xi_1}}\otimes\ket{0}+\ket{0}\otimes\ket{\vec{\xi_2}}\bigg).
\end{equation}
In the density matrix form we can write $\rho_m(t_i)=\ket{\Psi(t_i)}_m\prescript{}{m}{\bra{\Psi(t_i)}}$.

\begin{figure*}
	\centering 
	\includegraphics[width=12cm]{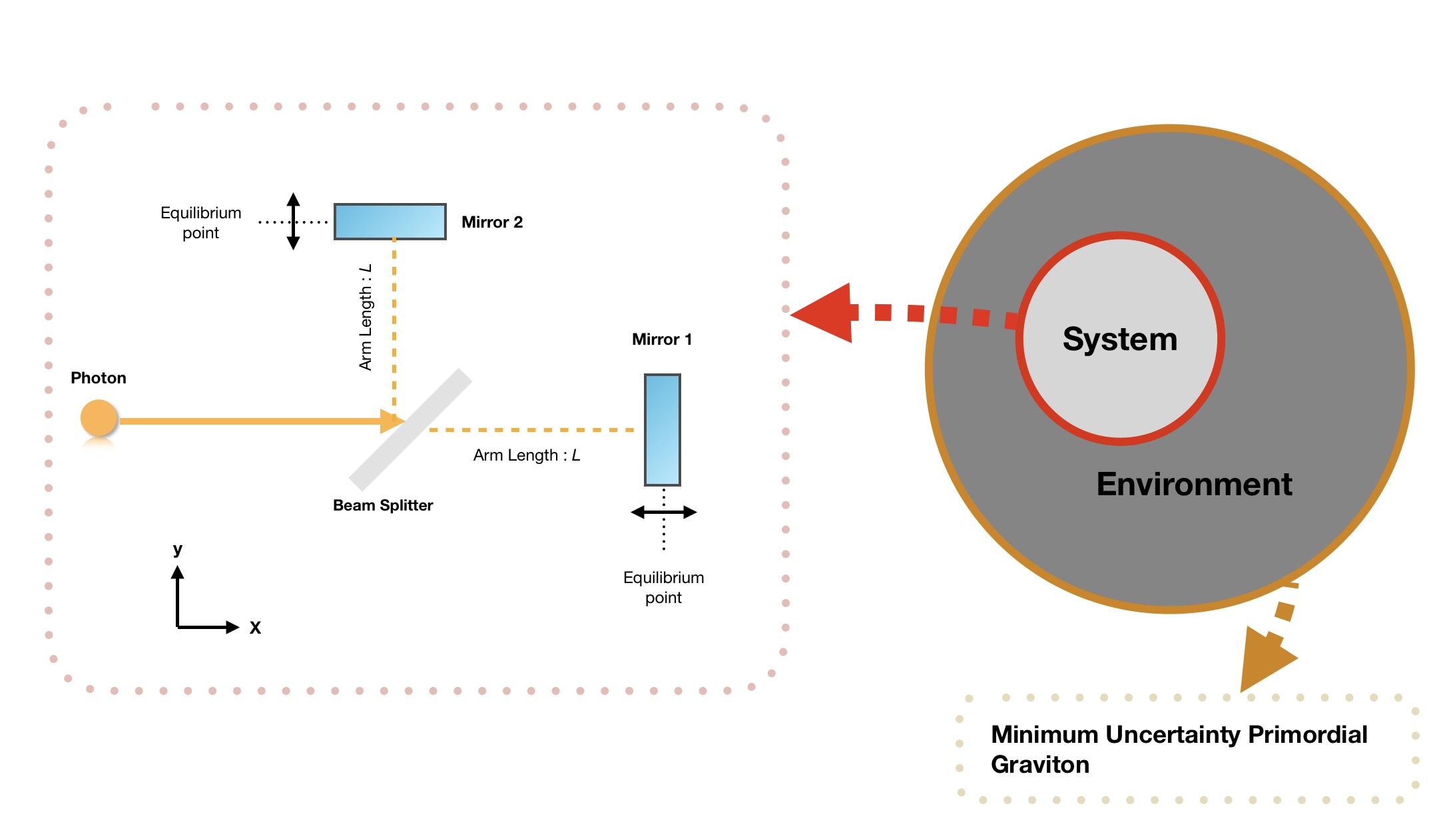}
	\caption{\label{fig2} The Michelson equal arm interferometer (the experimental setup $(\rho_m)$) in the environment filled with graviton originated from the early universe with the minimum uncertainty initial state $(\rho^{\otimes})$. }
\end{figure*}

There will be primordial gravitational waves in its environment with an initial minimum uncertainty state, as shown in Figure (\ref{fig2}). This means the quantum state will be the equation (\ref{eq:superposition1}) or (\ref{eq:superposition2}). Because we want to study the influence of the presence of non-classical correlations in the initial state of the primordial graviton on its decoherence, we assume that gravitational waves in the environment will only have one polarization mode (in this study, is chosen to be the polarization $\otimes$).Therefore, the $\oplus$ mode in the quantum state equations (\ref{eq:superposition1}) and (\ref{eq:superposition2}) will be traced out. In density matrix representation, it can be written as follows
\begin{eqnarray}\label{eq:Dens1}
\rho^{\otimes}&=&Tr_{\oplus}\Big(\ket{\psi}\bra{\psi}\Big)\nonumber\\
&=&\tanh^2 u \ket{0^{\otimes}}{\bra{0^{\otimes}}}+(1-\tanh^2 u)\,\,\sum_{n=0}^{N}\,\,\frac{1}{2^{N}}\,\,\begin{pmatrix}
N \\
n
\end{pmatrix}\ket{n^{\otimes}_{\mathbf{k}}}\bra{n^{\otimes}_{\mathbf{k}}}+\tanh u\,\,(1-\tanh^2 u)^{1/2}\nonumber\\
&\,\,&\times\frac{1}{2^{N/2}}\,\,\Big[\ket{0^{\otimes}}{\bra{N^{\otimes}_{\mathbf{k}}}}+\ket{N^{\otimes}_{\mathbf{k}}}{\bra{0^{\otimes}}}\Big]\nonumber\\
&=&\rho^{\otimes}_{Diag}+\rho^{\otimes}_{Non-Diag}.
\end{eqnarray}
Both equations (\ref{eq:superposition1}) and (\ref{eq:superposition2}) will have the same form. In general, the density matrix $\rho^{\otimes}$ can be separated based on diagonal and non-diagonal elements, with
\begin{equation}\label{eq:5.4}
\rho^{\otimes}_{Diag}=\tanh^2 u \ket{0^{\otimes}}{\bra{0^{\otimes}}}+(1-\tanh^2 u)\,\,\sum_{n=0}^{N}\,\,\frac{1}{2^{N}}\,\,\begin{pmatrix}
N \\
n
\end{pmatrix}\ket{n^{\otimes}_{\mathbf{k}}}\bra{n^{\otimes}_{\mathbf{k}}}
\end{equation}
and
\begin{equation}\label{eq:5.5}
\rho^{\otimes}_{Non-Diag}=\tanh u\,\,(1-\tanh^2 u)^{1/2}\frac{1}{2^{N/2}}\,\,\Big[\ket{0^{\otimes}}{\bra{N^{\otimes}_{\mathbf{k}}}}+\ket{N^{\otimes}_{\mathbf{k}}}{\bra{0^{\otimes}}}\Big].
\end{equation}
By assuming that the system's environment has only one type of polarization, the non-diagonal elements of the environmental density matrix will have a simple form such as equation (\ref{eq:5.5}). The total density matrix of the system and environment can be written as follows
\begin{eqnarray}
\rho(t_i)&=&\rho_m(t_i)\otimes\rho^{\otimes}\nonumber\\
&=&\Big(\rho_{11}(t_i)+\rho_{22}(t_i)+\rho_{12}(t_i)+\rho_{21}(t_i)\Big)\otimes\Big(\rho^{\otimes}_{Diag}+\rho^{\otimes}_{Non-Diag}\Big),
\end{eqnarray}

The density matrix of the system and environment evolves over the time based on the Langevin equation of geodesic deviation of the mirror in the presence of gravitons. Due to the interaction between the mirror and gravitons, the quantum state of the system and the environment will be entangled. This means to obtain the quantum state of the mirror at a certain time, we will trace out the degrees of freedom of the graviton environment from the total density matrix as in Appendix (\ref{ApenB}). We will obtain the density matrix of the mirror at a certain time induced by gravitons (for example, we chose the maximum number of initial gravitons $N=3$ and $\rho_f$ to be equation (\ref{eq:Dens1})) as
\begin{eqnarray}\label{eq:DensFinal}
\rho_m(t_f)&\approx&\rho_{11}(t_i)+\rho_{22}(t_i)+\Big(\exp\Big\{\Phi_{Diag\,\,N=3}[\xi_1,\xi_2]\Big\}+\exp\Big\{\Phi_{Non-Diag\,\,N=3}[\xi_1,\xi_2]\Big\}\Big)\,\,\rho_{12}(t_i)\nonumber\\
&\,\,&+\Big(\exp\Big\{\Phi^*_{Diag\,\,N=3}[\xi_1,\xi_2]\Big\}+\exp\Big\{\Phi^*_{Non-Diag\,\,N=3}[\xi_1,\xi_2]\Big\}\Big)\,\,\rho_{21}(t_i)
\end{eqnarray}
The influence of the environmental gravitons will only affect the interference terms of the mirrors. However, unlike research \cite{Kanno4}, the influence phase functional in this case will be separated into two parts based on the diagonal $\big(\Phi_{Diag}[\xi_1,\xi_2]\big)$ and non-diagonal $\big(\Phi_{Non-Diag}[\xi_1,\xi_2]\big)$ terms of the environmental density matrix. The real part of the influence phase functional will be a factor that influences the existence of the interference terms of the matter density matrix. Because the real part is time-dependent, at a certain time the interference terms will disappear. In this condition, the entanglement state of the interferometer mirrors will be destroyed. This process of losing the interference terms is known as decoherence. Which is the real part of the influence phase functional is called the decoherence functional and the time when the interference term is zero is called the decoherence time. In this study there will be two influence phase functional as shown in equation (\ref{eq:DensFinal}), meaning that there will also be two decoherences functional, each of which will have the form
\begin{eqnarray}\label{eq:DecFunc1}
\Gamma_{Diag}(t_f)=\frac{m^2}{8}\int^{t_f}_0 dt\int^{t_f}_0 dt'\,\,Tr\Big(\hat{S}^{\otimes\dagger}(\zeta)\big\{ \hat{N}_{ij}(t), \hat{N}_{kl}(t')\big\}\hat{S}^{\otimes}(\zeta)\,\,\rho^{\otimes}_{Diag}\Big)\,\,\Delta(\xi^i\xi^j)(t)\,\,\Delta(\xi^k\xi^l)(t')\nonumber\\
\end{eqnarray}
and
\begin{eqnarray}\label{eq:DecFunc2}
\Gamma_{Non-Diag\,\,N=3}(t_f)&=&\ln\Bigg[\,\,\bigg|\bigg(\frac{m^3}{24}\int^{t_f}_0 dt\int^{t_f}_0 dt'\int^{t_f}_0 dt''\,\,Tr\bigg(\hat{S}^{\otimes\dagger}(\zeta)\Big\{\hat{N}_{ij}(t),\big\{\hat{N}_{kl}(t'),\hat{N}_{mn}(t'')\big\}\Big\}\nonumber\\
&\,\,&\times\hat{S}^{\otimes}(\zeta)\,\,\rho^{\otimes}_{Non-Diag\,\,N=3}\bigg)\Delta(\xi^i\xi^j)(t)\,\,\Delta(\xi^k\xi^l)(t')\,\,\Delta(\xi^m\xi^n)(t'')\bigg)^{-1}\bigg|\,\,\Bigg]\nonumber\\
&\equiv&\ln\Big[\,\,\big|\mathcal{N}(t_f)\big|\Big]
\end{eqnarray}
Here $\Delta(\xi^i\xi^j)(t)=\xi^i_1(t)\xi^j_1(t)-\xi^i_2(t)\xi^j_2(t)$, that will be determined based on the experimental setup in Figure (\ref{fig2}). Equation (\ref{eq:DecFunc1}) is the decoherence functional that arises due to the influence of the diagonal terms of the environmental density matrix. This expression will apply not only to the diagonal density matrices of equations (\ref{eq:Dens1}) for $N=3$ but to all environmental density matrices of primordial gravitons with only diagonal elements. It is different from equation (\ref{eq:DecFunc2}), which this expression specifically appears for the non-diagonal density matrices (\ref{eq:Dens1}) with the maximum number of initial gravitons $N=3$. 

The existence of the squeezing operator on both decoherence functionals is due to gravitons in the environment originating from the early universe. Where squeezed formalism will be used, as in research \cite{Maity, Vennin}. This formalism is based on the cosmological inflation model, in which the initial quantum state will experience squeezing as the universe expands rapidly. Mathematically, the squeezing process will be represented as a two-mode squeezed operator as follows
\begin{equation}\label{squeezed}
\hat{S}^{A}(\zeta)=\exp\bigg[\zeta_k^*\,\,\hat{a}^A_{\textbf{k}}\hat{a}^A_{-\textbf{k}}-\zeta_k\,\,\hat{a}^{A\dagger}_{\textbf{k}}\hat{a}^{A\dagger}_{-\textbf{k}}\bigg],
\end{equation}
with $\zeta_k\equiv r_ke^{i\varphi_k}$, where $r_k$ and $\varphi_k$ is a squeezing and phase angle parameter. In general, the squeezing parameter $r_k$ and the phase $\varphi_k$ depend on $k$. However, for simplicity, we regard these variables as constant so that the squeezing parameter can have a value of $0\leq r_k$ and the phase angle parameter of $0\leq \varphi_k \leq 2 \pi$.  For the initial state of the Bunch-Davies vacuum $(\ket{0^A})$, due to the squeezing process, the vacuum state in the era of radiation domination $(\ket{0^A}_R)$ can be viewed as a state of entanglement as
\begin{eqnarray}
\ket{0^A}_R&=&\hat{S}^A(\zeta)\ket{0^A}\nonumber\\
&=&\frac{1}{\cosh r_k}\sum_{n=0}^{\infty}(-1)^n\,\,e^{in\varphi_k}\,\,\tanh^n r_k\,\,\ket{n^A_{\mathbf{k}},n^A_{-\mathbf{k}}}
\end{eqnarray}
This state is the entanglement between the wave number modes $\mathbf{k}$ and $-\mathbf{k}$.  It should be noted that the squeezed parameter $r_k$ here is different from the squeeze parameter in the initial state $(\lambda)$ that defines the degree of freedom of the deviation between the variances of gravitational wave intensity measurements with different types of polarization.

\subsection{Entanglement Initial State of Primordial Graviton}\label{Sec5.1}
In this section, we will calculate the decoherence functional for the initial state of the graviton environment in which the entanglement state (\ref{eq:Entang1}), and (\ref{eq:Entang2}) is chosen. If expressed in the form of a density matrix with one polarization mode (only mode $\otimes$), the entanglement state will take the form of the equation (\ref{eq:5.4}) with $\tanh u=0$, or
\begin{equation}\label{eq:DensEntang}
\rho^{\otimes}=\sum_{n=0}^{N}\,\,\frac{1}{2^{N}}\,\,\begin{pmatrix}
N \\
n
\end{pmatrix}\ket{n^{\otimes}_{\mathbf{k}}}\bra{n^{\otimes}_{\mathbf{k}}}.
\end{equation}
This density matrix only has diagonal elements. It means in the density matrix equation (\ref{eq:DensFinal}), $\exp\big\{\Phi_{Non-Diag\,\, N=3}[\xi_1,\xi_2]\big\}\approx0$. So, there will only be one functional decoherence, namely equation (\ref{eq:DecFunc1}). To calculate the decoherence functional, we will first calculate the two-point anticommutation relation $Tr\Big(\hat{S}^{\otimes\dagger}(\zeta)\big\{\hat{N}_{ij}(t), \hat {N}_{kl}(t')\big\}\hat{S}^{\otimes}(\zeta)\rho^{\otimes}\Big)$. By using the definition of the noise operator from the graviton in equation (\ref{Noise}), then
\begin{eqnarray}\label{eq:twopoint}
Tr\Big(\hat{S}^{\otimes\dagger}(\zeta)\big\{\hat{N}_{ij}(t), \hat{N}_{kl}(t')\big\}\hat{S}^{\otimes}(\zeta)\rho^{\otimes}\Big)&=&\frac{\kappa^2}{10\pi^2}\bigg(\delta_{ik}\delta_{jl}+\delta_{il}\delta_{jk}-\frac{2}{3}\delta_{ij}\delta_{kl}\bigg)\int_{0}^{\Omega_m}\,dk\,\,k^6 P(t,t',k),\nonumber\\
\end{eqnarray}
where $P(t,t',k)$, is
\begin{eqnarray}
Tr\bigg(\hat{S}^{\otimes\dagger}(\zeta)\big\{\hat{h}^{\otimes}_{\textbf{k},I}(t),\hat{h}^{\otimes}_{\textbf{k}',I}(t')\big\}\hat{S}^{\otimes}(\zeta)\rho^{\otimes}\bigg)=\delta_{\textbf{k}+\textbf{k}',0}P(t,t',k).
\end{eqnarray}
By using the density matrix (\ref{eq:DensEntang}) and squeezed operator (\ref{squeezed}), we got
\begin{eqnarray}\label{eq:P}
P(t,t',k)&=&\frac{(N+1)}{2k}\Big[\cos (k(t-t'))\cosh 2r_k-\cos (k(t-t')-\varphi_k)\sinh 2r_k\Big]
\end{eqnarray}

Based on the conventional inflation scenario, $\sinh 2r_k\simeq\cosh 2r_k\simeq(k_c/k)^4$, where $k_c=2\pi f_c$ and $f_c$ is the cutoff frequency of primordial gravitational waves. The bound on the cutoff frequency from CMB is $f_c\lesssim 10^9 \,\,$ Hz. Obtained
\begin{align}\label{eq:5.15}
\int_{0}^{\Omega_m}\,dk\,\,k^6 P(t,t',k)\approx\Big(N+1\Big)\,\,k^4_c\,\,\Omega^2_m\,\,\frac{x\sin(x)+\cos(x)-1}{x^2},
\end{align}
with $x=\Omega_m(t-t')$. Based on the experimental setup in Figure (\ref{fig2}), there will be two possible forms of $\Delta(\xi^i\xi^j)(t)$ that are
\begin{eqnarray}
\Delta\xi^2_1(t)&=&(L+A\cos \omega t)^2-L^2,\label{eq:5.16}\\
\Delta\xi^2_2(t)&=&L^2-(L+A\cos \omega t)^2.\label{eq:5.17}
\end{eqnarray}
If the oscillation amplitude is much smaller than the length of the interferometer arm, equation (\ref{eq:5.16}) and (\ref{eq:5.17}) can be approximated to be $\Delta\xi^2_i\sim2LA$. Substituting equation (\ref{eq:5.15}) into equation (\ref{eq:DecFunc1}) and using the same method as in the research \cite{Kanno4}, then the decoherence functional becomes
\begin{equation}\label{eq:5.18}
\Gamma_{Diag}(t_f)\approx\Big(N+1\Big)\,\,\frac{4\pi^3}{5}\bigg(\frac{m}{M_p}\bigg)^2(L\,\,f_c)^4\bigg(\frac{A}{L}\bigg)^2(\omega\,\,t_f)
\end{equation}

This decoherence functional is a general form of decoherence functional for the environment in the form of a Bunch-Davies vacuum. Where there will be a factor $(N+1)$, the role of this factor can be seen when we consider the decoherence time $t_f$. For $N=0$, the decoherence time (the condition when $\Gamma_{Diag}(t_f)\simeq1$) will be 20 seconds when the parameters $\omega=10^3$ Hz, $L=40$ km, $m=40$ kg, $f_c=10^9$ Hz, and $A=10/\sqrt{2m\omega}$ are selected. This means when $N\neq0$, the bigger the number of $N$, the decoherence time will be reduced to $20/(N+1)$ seconds. However, if the decoherence time is expected to remain around 20 seconds, this can be obtained by reducing the interferometer arm length $(L)$ or mirror mass $(m)$ to $(N+1)^{-1/2}$ times. This means that the entanglement state between the mirrors will last for about 20 seconds, the same as the initial state of the Bunch-Davies vacuum, which can also be used as evidence of the existence of primordial gravitons. So, this method of detecting primordial gravitons using decoherence can still be used if the initial state is in the form of equations (\ref{eq:Entang1}) and (\ref{eq:Entang2}). This initial state is a state that has non-classical correlation but does not have non-diagonal elements when looking at its density matrix for one of the polarization modes. 

\subsection{Superposition Initial State of Primordial Graviton}\label{Sec5.2}
This section will calculate the decoherence functional for the environmental density matrix as an equation (\ref{eq:Dens1}) for $N=3$. This density matrix will induce two decoherence functionals. Where the interference terms of the mirror density matrix equation (\ref{eq:DensFinal}) will not be zero if one of its decoherence functional values is small ($\Gamma(t_f)< 1$). For the diagonal decoherence functional $(\Gamma_{Diag}(t_f))$, the expression is not much different from equation \ref{eq:5.18} for $N=3$ (just replacing the factor $(N+1)$ with $(4-3\tanh^2 u)$), meaning that the decoherence time of this function will only last for a maximum of 20 seconds. Meanwhile, for the non-diagonal decoherence functional $(\Gamma_{Non-Diag\,\,N=3}(t_f))$, different results will be obtained.

To start the discussion, we will calculate the three-point correlation from the non-diagonal decoherence  functional of equation (\ref{eq:DecFunc2}), which will take the following form
\begin{align}\label{eq:threepoint}
Tr\bigg(\hat{S}^{\otimes\dagger}(\zeta)\Big\{\hat{N}_{ij}(t),\big\{\hat{N}_{kl}(t'),\hat{N}_{mn}(t'')\big\}&\Big\}\hat{S}^{\otimes}(\zeta)\,\,\rho^{\otimes}_{Non-Diag\,\,N=3}\bigg)\nonumber\\
&=\frac{\bigg(\delta_{ik}\delta_{jl}+\delta_{il}\delta_{jk}-\frac{2}{3}\delta_{ij}\delta_{kl}\bigg)}{10\pi^2M^3_{p}}\,\,\int_{0}^{\Omega_m}dk\,\,k^8\,\,e^{\otimes}_{\textbf{k},mn}\,\,F(t,t',t'',k),
\end{align}
with $F(t,t',t'',k)$ is
\begin{align}
Tr\bigg(\hat{S}^{\otimes\dagger}(\zeta)\Big\{\delta\hat{h}^{\otimes}_{\textbf{k},I}(t),\big\{ \delta\hat{h}^{\otimes}_{\textbf{k}',I}(t'),\delta\hat{h}^{\otimes}_{\textbf{k}'',I}(t'')\big\}\Big\}\hat{S}^{\otimes}(\zeta)\,\,\rho^{\otimes}_{Non-Diag\,\,N=3}\bigg)
=\delta_{\textbf{k}-\textbf{k}',0}\,\,\delta_{\textbf{k}'+\textbf{k}'',0}\,\,F\,\,(t,t',t'',k)
\end{align}
and the polarisation tensor $e^{\otimes}_{\textbf{k},mn}$ will look like \cite{Weinberg}
\begin{equation}
e^A_{\textbf{k},mn}=\frac{1}{\sqrt{2}}\begin{pmatrix}
1 & \mp i & 0 \\
\mp i & 1 & 0 \\
0 & 0 & 0
\end{pmatrix}.
\end{equation}
The $\pm$ sign in the matrix refers to the choice of the polarization mode of the primordial gravitational wave. Furthermore, because $\Delta(\xi^i\xi^j)(t)$ is only non-zero when the index $i=j$ as shown in equations (\ref{eq:5.16}) and (\ref{eq:5.17}), the part of the polarization tensor that affects the calculation of the three-point correlation function is only the diagonal elements. Taking that $\cosh 3r_k\sim(k_c/k)^6$ we got
\begin{eqnarray}
F(t,t',t'',k)&\approx&\frac{\sqrt{6}}{12\,\,k^{3/2}}\,\,\tanh u\big(1-\tanh^2 u\big)^{1/2}\bigg[\bigg(\frac{k_c}{k}\bigg)^6\,\,\cos\Big(k\big(t+t'+t''\big)\Big)\bigg]
\end{eqnarray}
then
\begin{equation}
\int_{0}^{\Omega_m}dk\,\,k^8\,\,e^{\otimes}_{\textbf{k},mn}\,\,F(t,t',t'',k)\approx\frac{\sqrt{6}}{24}\,\,k^6_c\,\,\tanh u(1-\tanh^2 u)^{1/2}\,\,\frac{\Omega^{1/2}_m\,\,\sin\big(\Omega(t+t'+t'')\big)}{(t+t'+t'')}.
\end{equation}
Substituting into equation (\ref{eq:threepoint}) then equation (\ref{eq:DecFunc2}), using some integration and choosing the UV cutoff $\Omega_m\sim\xi^{-1}\equiv L^{-1}$ we find 
\begin{eqnarray}
\mathcal{N}(t_f)^{-1}&\approx&\frac{3\sqrt{3}m^2\pi^4}{5M^3_p}\big(A\,\,f^2_c\big)^3\,\,\tanh u\big(1-\tanh^2 u\big)^{1/2}\bigg(\frac{(t_f\,\,L^{-1/2})-2L^{3/2}\,\,(4\omega^2\,\,t^2_f+2)}{8\omega^2L^2-1}\bigg).\nonumber\\
\end{eqnarray}

The non-diagonal decoherence functional can be obtained by calculating the natural logarithm of the absolute value of $\mathcal{N}(t_f)$. If the same parameters are taken as in section (\ref{Sec5.1}), then the decoherence time is in the order $\sim\big(\tanh u\sqrt{1-\tanh^2 u}\,\,\big)^{-1/2} 10^9$ second. This result is much longer than the decoherence time from the diagonal decoherence functional. Unless the constant $u\sim 0$ or $\tanh u\sim 1$ (when the non-diagonal density matrix $(\rho^{\otimes}_{Non-Diag})$ vanishes, this condition is met). Similar results will also occur if the maximum number of initial gravitons is other than three. This means that the interference terms of the mirror density matrix will last much longer because of the non-diagonal terms of the environment density matrix.  The duration of the decoherence time, or the entanglement between the mirrors, that is too long will make it difficult for us to prove that there has been an interaction between the mirrors and the primordial graviton because the very long decoherence time makes it seem as if the primordial graviton is unable to break the entanglement of the mirror. So, if the initial state is the equations (\ref{eq:superposition1}) and (\ref{eq:superposition2}) (has non-diagonal elements in one of the polarization modes of its density matrix), this method is ineffective for detecting gravitons.

\section{Conclusion}\label{sec6}
We have investigated the decoherence induced by the primordial graviton, using the influence functional method, to show whether this method is still effective in detecting graviton if the initial state is not a Bunch-Davies vacuum but rather a minimum uncertainty state. To obtain an initial state with minimum uncertainty conditions, we use the operators that can define the polarization intensity of gravitational waves, namely the Stokes operator. It is found that the initial state of the primordial graviton can be an entanglement state between the polarization modes (equation (\ref{eq:Entang1}) or (\ref{eq:Entang2})), or in a more general form, it can be a superposition state between the vacuum and that entanglement (equation (\ref{eq:superposition1}) and (\ref{eq:superposition2})). In one of the polarization modes, the density matrix of the superposition states will have diagonal and non-diagonal elements. Meanwhile, the entanglement state will only have diagonal elements. Based on quantum discord calculations, these two states will have a non-classical correlation between the two polarization modes, which the Bunch-Davies vacuum does not. The non-classical correlation will increase if the total graviton increases.

To investigated the decoherence, we have first looked for the expression of the decoherence functional if the experimental setup of the system is in the same form as S. Kanno \textit{et al.}'s research \cite{Kanno4}. Where a Michelson equal arm interferometer, which has two macroscopic suspended mirrors at the end of each interferometer arm, is used. The experimental setup will be in an environment containing gravitons from the early universe. Using the influence functional method and because the initial state of the primordial graviton can have non-diagonal elements, there will be two decoherence functional caused by the diagonal and non-diagonal elements of the initial graviton density matrix. To calculate each decoherence functional, squeezed formalism will be used.

For the initial state of the primordial graviton in the form of an entanglement state between polarization modes equations (\ref{eq:Entang1}) or (\ref{eq:Entang2}), where the density matrix expression of this state for one of the polarization modes will not have non-diagonal elements, the maximum decoherence time is about 20 seconds. That can be fulfilled if the length of the interferometer arm and mirror mass in the experimental setup is $(N+1)^{-1/2}$ times the interferometer arm and mirror mass used in the initial state of the Bunch-Davies vacuum to obtain the decoherence time for about 20 seconds. It means that if the maximum number of initial gravitons $(N)$ increases, the dimensions of the experimental setup will be smaller, and because the decoherence time is short for this initial state of primordial graviton, the graviton detection using the research method by S. Kanno \textit{et al.} \cite{Kanno4} can still be used. However, different results will be obtained if the initial state of the primordial graviton is chosen to be in the form of the superposition equations (\ref{eq:superposition1}) and (\ref{eq:superposition2}). The presence of non-diagonal elements in the density matrix form for one of the polarization modes of this state will induce an additional decoherence functional. Consequently, the decoherence time can last much longer than the vacuum or entanglement initial states of the primordial graviton, which will be maintained for about $10^{9}$ seconds for $N = 3$. A similar result would also occur if the total number of graviton is other than three. As stated in the introduction section, a decoherence time that is too long will make it difficult to prove that there has been an interaction between the mirrors and the primordial graviton, because the very long decoherence time makes it seem as if the primordial graviton is unable to break the entanglement of the mirror. So, the graviton detection using the research method by S. Kanno \textit{et al.} \cite{Kanno4} can still be used if the initial state of the primordial graviton is a state with a non-classical correlation whose density matrix does not have non-diagonal elements.

\section*{Acknowledgement}
F.P.Z. and G.H. would like to thank Kemenristek DIKTI Indonesia for financial
supports. A.T. would like to thank the members of Theoretical
Physics Groups of Institut Teknologi Bandung for the hospitality.

\appendix

\section{The Projected State}\label{ApenA}
In this appendix, it will be shown that the minimum value of the entropy $S_{\otimes|\Pi^{\oplus}}(\rho^{\oplus\otimes})=0$, by showing that the projected state $\rho^{\otimes}_j$ is a pure state if $\rho^{\oplus\otimes}$ is also a pure state. Consider the general form of the pure bipartite state between the $\otimes$ and $\oplus$ polarization modes as follows
\begin{equation}
\ket{\psi}=\sum_{n,m}\,\,c_{nm}\ket{n^{\oplus}_{\mathbf{k}}, m^{\otimes}_{\mathbf{k}}},
\end{equation}
with $\sum_{n,m}\,\,c_{nm}\,c^*_{nm}=1$. In density matrix form, it can be written as
\begin{eqnarray}
\rho^{\oplus\otimes}&=&\ket{\psi}\bra{\psi}\nonumber\\
&=&\sum_{n,m}\sum_{p,q}\,\,c_{nm}\,c^*_{pq}\ket{n^{\oplus}_{\mathbf{k}}}\bra{p^{\oplus}_{\mathbf{k}}}\otimes\ket{m^{\otimes}_{\mathbf{k}}}\bra{q^{\otimes}_{\mathbf{k}}}
\end{eqnarray}
Then, to calculate the projected state $\rho^{\otimes}_j=\frac{1}{P_j}\,\,Tr_{\oplus}\bigg(\Big(\Pi^{\oplus}_j\otimes I\Big)\rho^{\oplus\otimes}\Big(\Pi^{\oplus}_j\otimes I\Big)\bigg)$, the projection operator is chosen to be $\Pi^{\oplus}_j=\ket{j^{\oplus}_{\mathbf{k}}}\bra{j^{\oplus}_{\mathbf{k}}}$, we got
\begin{eqnarray}
\Big(\Pi^{\oplus}_j\otimes I\Big)\rho^{\oplus\otimes}\Big(\Pi^{\oplus}_j\otimes I\Big)&=&\sum_{nm}\sum_{pq}\,\,c_{nm}c^*_{pq}\Big(\ket{j^{\oplus}_{\mathbf{k}}}\bra{j^{\oplus}_{\mathbf{k}}}\otimes I\Big)\rho^{\oplus\otimes}\Big(\ket{j^{\oplus}_{\mathbf{k}}}\bra{j^{\oplus}_{\mathbf{k}}}\otimes I\Big)\nonumber\\
&=&\sum_{nm}\sum_{pq}\,\,c_{nm}c^*_{pq}\ket{j^{\oplus}_{\mathbf{k}}}\braket{j^{\oplus}_{\mathbf{k}}|n^{\oplus}_{\mathbf{k}}}\braket{p^{\oplus}_{\mathbf{k}}|j^{\oplus}_{\mathbf{k}}}\bra{j^{\oplus}_{\mathbf{k}}}\otimes\ket{m^{\otimes}_{\mathbf{k}}}\bra{q^{\otimes}_{\mathbf{k}}}\nonumber\\
&=&\sum_{m,q}\,\,c_{jm}\,c^*_{jq}\ket{j^{\oplus}_{\mathbf{k}}}\bra{j^{\oplus}_{\mathbf{k}}}\otimes\ket{m^{\otimes}_{\mathbf{k}}}\bra{q^{\otimes}_{\mathbf{k}}}
\end{eqnarray}
so that
\begin{eqnarray}
P_j&=&Tr\bigg(\Big(\Pi^{\oplus}_j\otimes I\Big)\rho^{\oplus\otimes}\Big(\Pi^{\oplus}_j\otimes I\Big)\bigg)\nonumber\\
&=&\sum_{kl}\Big(\bra{k^{\oplus}_{\mathbf{k}}}\otimes\bra{l^{\otimes}_{\mathbf{k}}}\Big)\,\,\sum_{m,q}\,\,c_{jm}\,c^*_{jq}\ket{j^{\oplus}_{\mathbf{k}}}\bra{j^{\oplus}_{\mathbf{k}}}\otimes\ket{m^{\otimes}_{\mathbf{k}}}\bra{q^{\otimes}_{\mathbf{k}}}\,\,\Big(\ket{k^{\oplus}_{\mathbf{k}}}\otimes\ket{l^{\otimes}_{\mathbf{k}}}\Big)\nonumber\\
&=&\sum_{kl}\sum_{m,q}\,\,c_{jm}\,c^*_{jq}\braket{j^{\oplus}_{\mathbf{k}}|k^{\oplus}_{\mathbf{k}}}\braket{k^{\oplus}_{\mathbf{k}}|j^{\oplus}_{\mathbf{k}}}\otimes\braket{l^{\oplus}_{\mathbf{k}}|m^{\oplus}_{\mathbf{k}}}\braket{q^{\oplus}_{\mathbf{k}}|l^{\oplus}_{\mathbf{k}}}\nonumber\\
&=&\sum_l\,\,c_{jl}c^*_{jl}
\end{eqnarray}
and
\begin{eqnarray}
\rho^{\otimes}_j&=&\frac{1}{P_j}\,\,Tr_{\oplus}\bigg(\Big(\Pi^{\oplus}_j\otimes I\Big)\rho^{\oplus\otimes}\Big(\Pi^{\oplus}_j\otimes I\Big)\bigg)\nonumber\\
&=&\frac{1}{\sum_l\,\,c_{jl}c^*_{jl}}\,\,\sum_{k}\Big(\bra{k^{\oplus}_{\mathbf{k}}}\otimes I\Big)\,\,\sum_{m,q}\,\,c_{jm}\,c^*_{jq}\ket{j^{\oplus}_{\mathbf{k}}}\bra{j^{\oplus}_{\mathbf{k}}}\otimes\ket{m^{\otimes}_{\mathbf{k}}}\bra{q^{\otimes}_{\mathbf{k}}}\,\,\Big(\ket{k^{\oplus}_{\mathbf{k}}}\otimes I\Big)\nonumber\\
&=&\frac{1}{\sum_l\,\,c_{jl}c^*_{jl}}\,\,\sum_{k}\sum_{m,q}\,\,c_{jm}\,c^*_{jq}\braket{j^{\oplus}_{\mathbf{k}}|k^{\oplus}_{\mathbf{k}}}\braket{k^{\oplus}_{\mathbf{k}}|j^{\oplus}_{\mathbf{k}}}\otimes\ket{m^{\otimes}_{\mathbf{k}}}\bra{q^{\otimes}_{\mathbf{k}}}\nonumber\\
&=&\frac{1}{\sum_l\,\,c_{jl}c^*_{jl}}\,\,\sum_{m,q}\,\,c_{jm}\,c^*_{jq}\ket{m^{\otimes}_{\mathbf{k}}}\bra{q^{\otimes}_{\mathbf{k}}}
\end{eqnarray}
This density matrix is a pure state. It will be proven by showing that $(\rho^{\otimes}_j)^2=\rho^{\otimes}_j$ as follow
\begin{eqnarray}
\big(\rho^{\otimes}_j\big)^2&=&\rho^{\otimes}_j\,\,\rho^{\otimes}_j\nonumber\\
&=&\Bigg(\frac{1}{\sum_l\,\,c_{jl}c^*_{jl}}\,\,\sum_{m,q}\,\,c_{jm}\,c^*_{jq}\ket{m^{\otimes}_{\mathbf{k}}}\bra{q^{\otimes}_{\mathbf{k}}}\Bigg)\Bigg(\frac{1}{\sum_k\,\,c_{jk}c^*_{jk}}\,\,\sum_{n,p}\,\,c_{jn}\,c^*_{jp}\ket{n^{\otimes}_{\mathbf{k}}}\bra{p^{\otimes}_{\mathbf{k}}}\Bigg)\nonumber\\
&=&\frac{1}{\sum_l\,\,c_{jl}c^*_{jl}\,\,\sum_k\,\,c_{jk}c^*_{jk}}\,\,\sum_{m,q}\sum_{n,p}\,\,c_{jm}\,c^*_{jq}c_{jn}\,c^*_{jp}\ket{m^{\otimes}_{\mathbf{k}}}\braket{q^{\oplus}_{\mathbf{k}}|n^{\oplus}_{\mathbf{k}}}\bra{p^{\otimes}_{\mathbf{k}}}\nonumber\\
&=&\frac{1}{\sum_l\,\,c_{jl}c^*_{jl}\,\,\sum_k\,\,c_{jk}c^*_{jk}}\,\,\sum_q\,\,c_{jq}c^*_{jq}\sum_{m,p}\,\,c_{jm}\,c^*_{jp}\ket{m^{\otimes}_{\mathbf{k}}}\bra{p^{\otimes}_{\mathbf{k}}}\nonumber\\
&=&\frac{1}{\sum_l\,\,c_{jl}c^*_{jl}}\,\,\sum_{m,p}\,\,c_{jm}\,c^*_{jp}\ket{m^{\otimes}_{\mathbf{k}}}\bra{p^{\otimes}_{\mathbf{k}}}\nonumber\\
&=&\rho^{\otimes}_j.
\end{eqnarray}
So that $S(\rho^{\otimes}_j)=0$ and $S_{\otimes|\Pi^{\oplus}}(\rho^{\oplus\otimes})=0$.

\section{Influence Functional Method}\label{ApenB}
This section of the appendix will explain the influence functional method, as in the research \cite{Breuer}, for determining the expression of the decoherence functional. However, in this explanation, the environmental density matrix will be assumed to be an equation (\ref{eq:Dens1}) with non-diagonal terms. This appendix will be divided into two parts. The B.1 part will explain the decoherence functional in the QED (quantum electrodynamics)  case, and the B.2 part is for the gravity case.

\subsection{Influence Functional in QED}
Before discussing the decoherence functional for the gravitational case, we first considered the decoherence functional with the same analogy (QED). Suppose a matter $(\rho_m)$ is in an environment with electromagnetic field radiation. In the initial state, the density matrix can be written as $\rho(t_i)=\rho_m(t_i)\otimes\rho_f(t_i)$, where $\rho_f(t)$ is the environmental density matrix. If we want to calculate the matter density matrix at a final time $\rho_m(t_f)$, then
\begin{equation}\label{eq:Trace}
\rho_m(t_f)=Tr_f\bigg(T_{\leftarrow}\exp\Big(\int_{t_i}^{t_f}d^4x\mathcal{L}(x)\Big)\rho(t_i)\bigg).
\end{equation}
Where $T_{\leftarrow}$ is the time-ordering operator and $\mathcal{L}(x)$ is the Liouville super operator, which has the relation
\begin{equation}
\mathcal{L}(x)\rho=-i[\mathcal{H}(x),\rho]
\end{equation}
$\mathcal{H}(x)$ denotes the Hamiltonian density at spacetime coordinates $x=x^{\mu}=(x^0,\vec{x})$. We choose the coulomb gauge in the following, which means that the Hamiltonian density takes the form $\mathcal{H}(x)=\mathcal{H}_C(x)+\mathcal{H}_{tr}(x)$, with
\begin{eqnarray}
\mathcal{H}_C(x)=\frac{1}{2}\int\,\,d^3y\frac{j^0(x^0,\vec{x})j^0(x^0,\vec{y})}{4\pi|\vec{x}-\vec{y}|},
\end{eqnarray}
and
\begin{eqnarray}
\mathcal{H}_{tr}(x)=j^{\mu}(x)A_{\mu}(x).
\end{eqnarray}
$\mathcal{H}_{tr}(x)$ represents the Hamiltonian density of the interaction of the matter current density $j^{\mu}(x)$ with the transverse radiation field $A^{\mu}( x)$. Next, the time-ordering operator $T_{\leftarrow}$ will be decomposed into a time-ordering operator $T_{\leftarrow}^j$ for matter current and a time-ordering operator $T_{\leftarrow}^A$ for electromagnetic fields. With some derivation and defined current super operator $J^{\mu}_{+}(x)\rho\equiv j^{\mu}(x)\rho$ and $J^{\mu}_{-} (x)\rho\equiv\rho j^{\mu}(x)$, then equation (\ref{eq:Trace}) becomes
\begin{eqnarray}\label{eq:trace2}
\rho_m(t_f)&=&T^j_{\leftarrow}\bigg(\exp\bigg[\int_{t_i}^{t_f}d^4x\mathcal{L}_C(x)-\frac{1}{2}\int_{t_i}^{t_f}d^4x\int_{t_i}^{t_f}d^4x'\theta(t-t')\,\,[A_{\mu}(x),A_{\nu}(x')] J^{\mu}_{+}(x)J^{\nu}_{+}(x')\nonumber\\
&\,\,&+\frac{1}{2}\int_{t_i}^{t_f}d^4x\int_{t_i}^{t_f}d^4x'\theta(t-t') [A_{\mu}(x),A_{\nu}(x')] J^{\mu}_{-}(x)J^{\nu}_{-}(x')\bigg] W[J_{+},J_{-}]\bigg).
\end{eqnarray}
Where
\begin{equation}\label{eq:Functional}
W[J_{+},J_{-}]=Tr_f\biggl\{\exp\bigg[\int_{t_i}^{t_f}d^4x \mathcal{L}_{tr}(x)\rho(t_i)\bigg]\biggr\}.
\end{equation}

This functional is part of the density matrix of matter at any time $(\rho_m(t_f))$ that contains the correlation between the matter field and electromagnetic radiation. In general, the initial radiation density matrix $\rho_f(t_i)$ will be chosen as a thermal equilibrium state where the density matrix of that state has only diagonal elements. In this study, the environmental state of the graviton is in conditions of minimum uncertainty that allow for non-diagonal elements in the density matrix. So $\rho_f(t_i)$ will be expanded and assumed to be the same as equation (\ref{eq:Dens1}). Furthermore, using an exponential expansion in equation (\ref{eq:Functional}) is obtained
\begin{eqnarray}
W[J_{+},J_{-}]&=&Tr_f\biggl\{\bigg[1+\int d^4x \mathcal{L}_{tr}(x)+\frac{1}{2!}\int d^4x\int d^4x'\mathcal{L}_{tr}(x)\mathcal{L}_{tr}(x')\nonumber\\
&\,\,&+\frac{1}{3!}\int d^4x\int d^4x'\int d^4 x''\mathcal{L}_{tr}(x)\mathcal{L}_{tr}(x')\mathcal{L}_{tr}(x'')+...\,\,\bigg]\rho_m(t_i)\otimes \rho_f(t_i)\biggr\}
\end{eqnarray}
The density matrix $\rho_f(t_i)$ in general can be split into a diagonal and a non-diagonal part $\Big(\rho_f(t)\equiv\rho_{f, Diag}(t)+\rho_{f, Non-Diag}(t)\Big)$. If the maximum number of initial gravitons $N$ is odd, then the diagonal density matrix $\rho_{f, Diag}(t)$ will only work on even-order Liouville super operators, while the non-diagonal density matrix $\rho_{f, Non-Diag}(t)$ will only work on odd-order Liouville super operators. This means that for $N=3$, the functional $W[J_{+},J_{-}]$ can be written as
\begin{eqnarray}\label{eq:Functional2}
W[J_{+},J_{-}]&=&\bigg(\exp\bigg[\frac{1}{2}\int d^4x\int d^4x'\,\,\,\,Tr_f\bigg\{\mathcal{L}_{tr}(x)\mathcal{L}_{tr}(x')\rho_{f,Diag\,\,N=3}(t_i)\bigg\}\bigg]\nonumber\\
&\,\,&+\frac{1}{3!}\int d^4x\int d^4x' \int d^4 x''Tr_f\biggl\{\mathcal{L}_{tr}(x)\mathcal{L}_{tr}(x')\mathcal{L}_{tr}(x'')\rho_{f,Non-Diag\,\,N=3}(t_i)\biggr\}\bigg)\rho_m(t_i).\nonumber\\
\end{eqnarray}
The Liouville super operators with odd orders are limited only to the third. The first term of the equation can be obtained by using the Wick theorem for the Gaussian correlation function, where the even-order correlation function can always be expressed as a two-point correlation function. From this result, if it is substituted back into equation (\ref{eq:trace2}),  and  defining a commutator superoperator $J_c(x)$
\begin{equation}
J^{\mu}_c(x)\rho\equiv[J^{\mu}_c(x),\rho]\,\,\,\,\,\,\,\,\,\,\,\,\,\,\,\,\,\,\,\,J^{\mu}_c(x)=J^{\mu}_+(x)-J^{\mu}_-(x)
\end{equation}
and an anticommutator superoperator $J_a(x)$
\begin{equation}
J^{\mu}_a(x)\rho\equiv \{J^{\mu}_c(x),\rho\}\,\,\,\,\,\,\,\,\,\,\,\,\,\,\,\,\,\,\,\,J^{\mu}_a(x)=J^{\mu}_+(x)+J^{\mu}_-(x),
\end{equation}
then we got
\begin{eqnarray}\label{eq:DensM}
\rho_m(t_f)&=&T^j_{\leftarrow}\Big(\exp\Big\{\Phi_{Diag\,\,N=3}[J_c,J_a]\Big\}+\exp\Big\{\Phi_{Non-Diag\,\,N=3}[J_c,J_a]\Big\}\Big)\rho_m(t_i).
\end{eqnarray}
$\Phi_{Diag\,\, N=3}[J_c, J_a]$ is the influence phase functional resulting from the diagonal density matrix $\Big(\rho_{f, Diag}(t)\Big)$ and $\Phi_{Non-Diag \,\, N=3}[J_c, J_a]$ is influence phase functional caused by the no-diagonal density matrix $\Big(\rho_{f, Non-Diag}(t)\Big)$. Each of those influence phase functional are
\begin{eqnarray}
\Phi_{Diag\,\,N=3}&=&\int_{t_i}^{t_f}d^4x\mathcal{L}_C(x)+\frac{1}{2}\int_{t_i}^{t_f}d^4x\int_{t_i}^{t_f}d^4x'\bigg[\,\,iD(x-x')_{\mu\nu}\,\,J^{\mu}_{c}(x)J^{\nu}_{a}(x')\nonumber\\
&\,\,&-Tr_f\Big(\big\{A_{\mu}(x),A_{\nu}(x')\big\}\,\,\rho_{f,Diag\,\,N=3}(t_i)\Big) J^{\mu}_{c}(x)J^{\nu}_{c}(x')\bigg]
\end{eqnarray}
and
\begin{eqnarray}
\Phi_{Non-Diag\,\,N=3}&=&\int_{t_i}^{t_f}d^4x\mathcal{L}_C(x)+\frac{1}{2}\int_{t_i}^{t_f}d^4x\int_{t_i}^{t_f}d^4x'\bigg[\,\,iD(x-x')_{\mu\nu}\,\,\Big(J^{\mu}_{c}(x)J^{\nu}_{a}(x')\nonumber\\
&\,\,&+J^{\mu}_{a}(x)J^{\nu}_{c}(x')\Big)\bigg]-\ln\Bigg[\,\,\bigg|\frac{1}{6}\int_{t_i}^{t_f}d^4x\int_{t_i}^{t_f}d^4x'\int_{t_i}^{t_f}d^4x''\,\,\nonumber\\
&\,\,&\times Tr_f\bigg(\Big\{A_{\mu}(x),\big\{A_{\nu}(x'),A_{\gamma}(x'')\big\}\Big\} \rho_{f,Non-Diag\,\,N=3}(t_i)\bigg)\nonumber\\
&\,\,&\times J^{\mu}_{c}(x)J^{\nu}_{c}(x')J^{\gamma}_{c}(x'')\bigg|^{-1}\bigg]\pm i\frac{\pi}{2},
\end{eqnarray}
with $D(x-x')_{\mu\nu}\equiv i[A_{\mu(x)},A_{\nu}(x')]$. The influence phase functional $\Phi_{Diag\,\, N=3}[J_c, J_a]$ is none other than the influence phase functional that is the same as obtained from the research of H. Breuer and F. Petruccione \cite{Breuer}. When the initial density matrix only has diagonal terms, $\Phi_{Non-Diag\,\, N=3}[J_c, J_a]$ will be zero. The last term in $\Phi_{Non-Diag\,\, N=3}[J_c, J_a]$ appears as a result of the complex logarithm when expressing the second term of equation (\ref{eq:Functional2}) in the form of an exponential. 
\begin{figure}[h]
	\centering 
	\includegraphics[width=9cm]{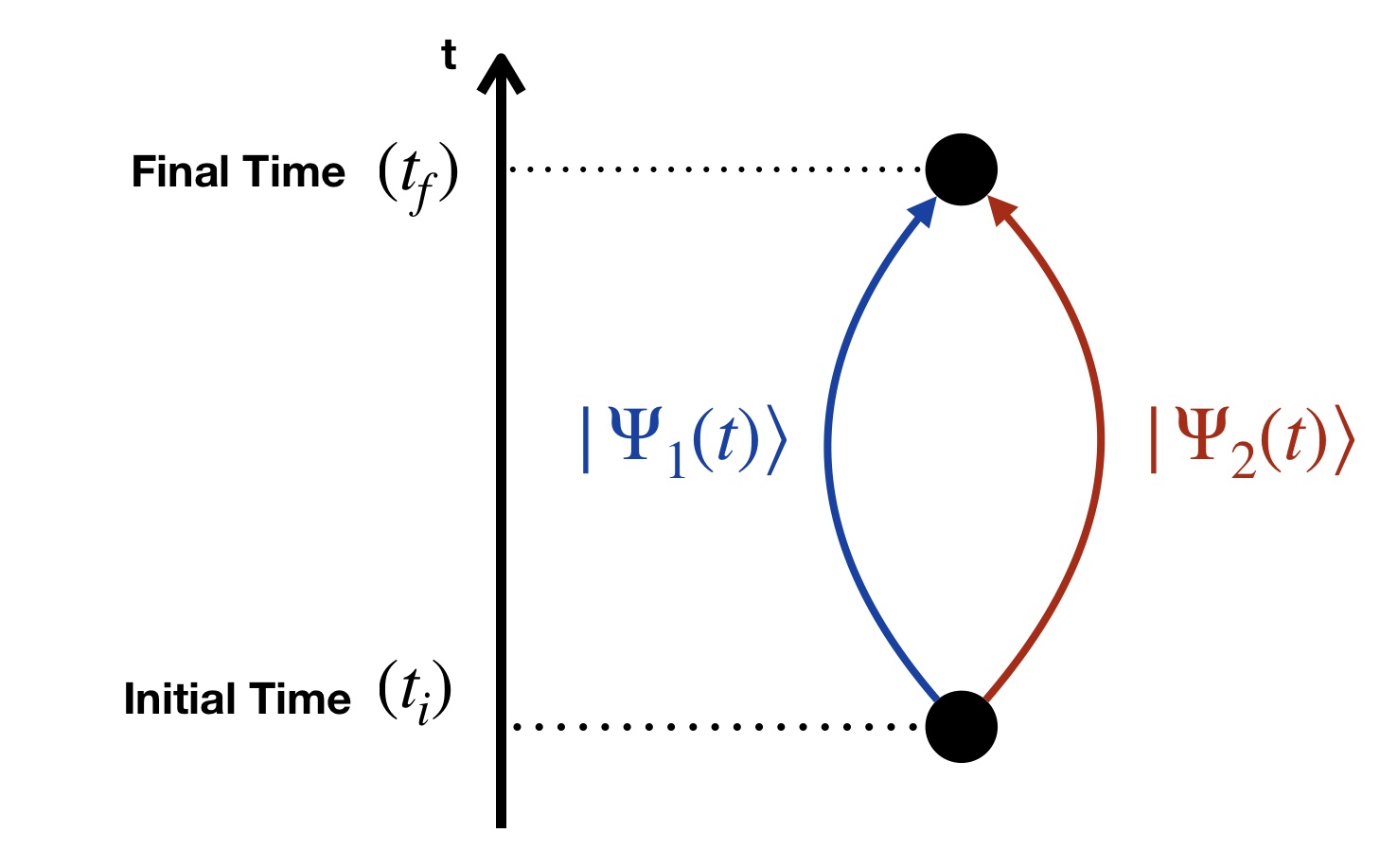}
	\caption{ \label{fig3} The trajectory of charge particle from initial time $(t_i)$ into some final time $(t_f)$.}
\end{figure}

The matter density matrix must be defined to find the degree of decoherence measurement. Assumed Figure (\ref{fig3}) describes the time-dependent quantum state of matter. Where a charged particle in the initial state can transfer to the final state through two trajectories, which will represent two amplitude probabilities (described by two wave packets, $\ket{\Psi_1(t)}$ and $\ket{\Psi_2(t)}$), then based on the superposition principle, the wave function at the initial time can be written as follows
\begin{equation}
\ket{\Psi(t_i)}_m=\ket{\Psi_1(t_i)}+\ket{\Psi_2(t_i)}.
\end{equation}
So the density matrix becomes
\begin{equation}\label{eq:DensM2}
\rho_m(t_i)=\rho_{11}(t_i)+\rho_{22}(t_i)+\rho_{12}(t_i)+\rho_{21}(t_i).
\end{equation}
If $j^{\mu}\ket{\Psi(t)}\approx s^{\mu}(x)\ket{\Psi(t)}$, with $s^{\mu}(x)$ is the classical current density then
\begin{eqnarray}
J^{\mu}_c(x)\,\,\rho_{11}\approx J^{\mu}_c(x)\,\,\rho_{22}\approx0,\,\,\,\,\,\,\,\,\,\,\,
J^{\mu}_c(x)\rho_{12}\approx\big[s^{\mu}_{1}(x)-s^{\mu}_{2}(x)\big]\,\,\,\,\,\,\,\,\,\,\,J^{\mu}_a(x)\rho_{12}\approx\big[s^{\mu}_{1}(x)+s^{\mu}_{2}(x)\big]\nonumber
\end{eqnarray}
and
\begin{equation}
\mathcal{L}_C(x)\,\,\rho_{11}\approx \mathcal{L}_C(x)\,\,\rho_{22}\approx0
\end{equation}
Substituted the matrix density equation (\ref{eq:DensM2}) into equation (\ref{eq:DensM}) is obtained
\begin{eqnarray}
\rho_m(t_f)&\approx&\rho_{11}(t_i)+\rho_{22}(t_i)+\Big(\exp\Big\{\Phi_{Diag\,\,N=3}[s_1,s_2]\Big\}+\exp\Big\{\Phi_{Non-Diag\,\,N=3}[s_1,s_2]\Big\}\Big)\,\,\rho_{12}(t_i)\nonumber\\
&\,\,&+\Big(\exp\Big\{\Phi^*_{Diag\,\,N=3}[s_1,s_2]\Big\}+\exp\Big\{\Phi^*_{Non-Diag\,\,N=3}[s_1,s_2]\Big\}\Big)\,\,\rho_{21}(t_i)
\end{eqnarray}
Where each influence phase functional now becomes
\begin{eqnarray}
\Phi_{Diag\,\,N=3}&=&\frac{1}{2}\int_{t_i}^{t_f}d^4x\int_{t_i}^{t_f}d^4x'\bigg[\,\,iD(x-x')_{\mu\nu}\,\,\big[s^{\mu}_{1}(x)-s^{\mu}_{2}(x)\big]\big[s^{\nu}_{1}(x')+s^{\nu}_{2}(x')\big]\nonumber\\
&\,\,&-Tr_f\Big(\big\{A_{\mu}(x),A_{\nu}(x')\big\}\,\,\rho_{f,Diag\,\,N=3}(t_i)\Big)\,\,\big[s^{\mu}_{1}(x)-s^{\mu}_{2}(x)\big]\big[s^{\nu}_{1}(x')-s^{\nu}_{2}(x')\big]\bigg]\nonumber\\
\end{eqnarray}
and
\begin{eqnarray}
\Phi_{Non-Diag\,\,N=3}&=&\frac{1}{2}\int_{t_i}^{t_f}d^4x\int_{t_i}^{t_f}d^4x'\Bigg[\,\,iD(x-x')_{\mu\nu}\,\,\bigg(\big[s^{\mu}_{1}(x)-s^{\mu}_{2}(x)\big]\big[s^{\nu}_{1}(x')+s^{\nu}_{2}(x')\big]\nonumber\\
&\,\,&+\big[s^{\mu}_{1}(x)+s^{\mu}_{2}(x)\big]\big[s^{\nu}_{1}(x')-s^{\nu}_{2}(x')\big]\bigg)\Bigg]-\ln\Bigg[\,\,\bigg|\frac{1}{6}\int_{t_i}^{t_f}d^4x\int_{t_i}^{t_f}d^4x'\int_{t_i}^{t_f}d^4x''\nonumber\\
&\,\,&Tr_f\bigg(\Big\{A_{\mu}(x),\big\{A_{\nu}(x'),A_{\gamma}(x'')\big\}\Big\}\,\,\rho_{f,Non-Diag\,\,N=3}(t_i)\bigg)\big[s^{\mu}_{1}(x)-s^{\mu}_{2}(x)\big]\nonumber\\
&\,\,&\times\big[s^{\nu}_{1}(x')-s^{\nu}_{2}(x')\big]\big[s^{\gamma}_{1}(x'')-s^{\gamma}_{2}(x'')\big]\bigg|^{-1}\,\,\Bigg]\pm\,\, i\frac{\pi}{2}.
\end{eqnarray}
These two influence phase functionals will only affect the interference terms of the density matrix of the matter. Where the real part of each influence phase functional could suppress the interference terms to be zero, which means there is information that will be lost due to environmental effects (decoherence). Therefore, the real part of those functionals can be defined as decoherence functional. In this case, two decoherence functional will arise as a consequence of the presence of diagonal and non-diagonal terms of the environmental density matrix. Those two decoherences functional can be expressed as follows 
\begin{eqnarray}\label{eq:DensFunc3}
\Gamma_{Diag}[\,\,\Delta s\,\,]=\frac{1}{2}\int_{t_i}^{t_f}d^4x\int_{t_i}^{t_f}d^4x'\,\,Tr_f\Big(\big\{A_{\mu}(x),A_{\nu}(x')\big\}\,\,\rho_{f,Diag}(t_i)\Big)\,\,\Delta s^{\mu}(x)\Delta s^{\nu}(x')
\end{eqnarray}
and
\begin{eqnarray}
\Gamma_{Non-Diag\,\,N=3}[\,\,\Delta s\,\,]&=&\ln\Bigg[\,\,\bigg|\frac{1}{6}\int_{t_i}^{t_f}d^4x\int_{t_i}^{t_f}d^4x'\int_{t_i}^{t_f}d^4x''\,\,Tr_f\bigg(\Big\{A_{\mu}(x),\big\{A_{\nu}(x'),A_{\gamma}(x'')\big\}\Big\}\nonumber\\
&\,\,&\times\rho_{f,Non-Diag\,\,N=3}(t_i)\bigg)\Delta s^{\mu}(x)\Delta s^{\nu}(x')\Delta s^{\gamma}(x'')\bigg|^{-1}\,\,\Bigg],
\end{eqnarray}
where $\Delta s^{\mu}(x)\equiv s^{\mu}_{1}(x)-s^{\mu}_{2}(x)$. For the decoherence functional equation (\ref{eq:DensFunc3}), because this equation would have the same form for all environment density matrices that only have diagonal elements, the notation $N=3$ is removed.
\subsection{Influence Functional in Gravity}
Next, we will look for the expression of decoherence functional for the gravity case as in section (\ref{sec4}). Where there are two test particles represented in the Fermi coordinates with a quantized gravitational waves environment (graviton). Formally, the expression of decoherence functional will be similar to the QED case. The time evolution that affects the correlation of the system with the environment is induced by a similar interaction Hamiltonian. In the case of gravity, the interaction Hamiltonian can be obtained from the last action term of the equation (\ref{eq:TotAction}). In the representation of the quantum noise, the interaction Hamiltonian can be expressed as
\begin{equation}
\hat{H}_{int}=-\frac{m}{2}\,\,\hat{N}_{ij}(t)\,\,\hat{\xi}^i(t)\hat{\xi}^j(t).
\end{equation}
The two decoherence functionals will be obtained using the same method as the QED case for $N = 3$ with the same environmental density matrix. Which could be written as 
\begin{eqnarray}
\Gamma_{Diag}(t_f)=\frac{m^2}{8}\int^{t_f}_0\,\,dt\int^{t_f}_0\,\,dt'\,\,Tr\Big(\big\{\hat{N}_{ij}(t),\hat{N}_{kl}(t)\big\}\,\,\rho^{\otimes}_{Diag}\Big)\,\,\Delta(\xi^i\xi^j)(t)\,\,\Delta(\xi^k\xi^l)(t')
\end{eqnarray}
and
\begin{eqnarray}
\Gamma_{Non-Diag\,\,N=3}(t_f)&=&\ln\Bigg[\,\,\bigg|\frac{m^3}{24}\int^{t_f}_0\,\,dt\int^{t_f}_0\,\,dt'\int^{t_f}_0\,\,dt''\,\,Tr\bigg(\Big\{\hat{N}_{ij}(t),\big\{\hat{N}_{kl}(t'),\hat{N}_{mn}(t'')\big\}\Big\}\nonumber\\
&\,\,&\times\rho^{\otimes}_{Non-Diag\,\,N=3}\bigg)\Delta(\xi^i\xi^j)(t)\,\,\Delta(\xi^k\xi^l)(t')\,\,\Delta(\xi^m\xi^n)(t'')\bigg|^{-1}\,\,\Bigg],
\end{eqnarray}
with $\Delta(\xi^i\xi^j)(t)=\xi^i_1(t)\xi^j_1(t)-\xi^i_2(t)\xi^j_2(t)$, that will be determined based on the experimental setup.

\end{document}